\documentclass[fleqn,usenatbib]{mnras}

\usepackage{newtxtext,newtxmath}
\usepackage{hyperref}
\usepackage{multirow}
\usepackage{tabularx}
\usepackage{xcolor}
\usepackage{bm}

\usepackage[T1]{fontenc}

\DeclareRobustCommand{\VAN}[3]{#2}
\let\VANthebibliography\thebibliography
\def\thebibliography{\DeclareRobustCommand{\VAN}[3]{##3}\VANthebibliography}

\usepackage{graphicx}	
\usepackage{amsmath}	

\newcommand{\tricolour}{\mbox{\textsc{Tricolour}}}
\newcommand{\rraturl}{\url{https://ratt.center/parrot}}
\newcommand{\wsclean}{\mbox{{\sc WSClean}}}
\newcommand{\breizorro}{\mbox{{\sc breizorro}}}
\newcommand{\cubical}{\mbox{{\sc CubiCal}}}
\newcommand{\quartical}{\mbox{{\sc QuartiCal}}}
\newcommand{\oldstimela}{\mbox{\textsc{Stimela}}}
\newcommand{\stimela}{\mbox{\textsc{Stimela2}}}
\newcommand{\caracal}{\mbox{\textsc{CARACal}}}
\newcommand{\smops}{\mbox{\textsc{smops}}}
\newcommand{\crystalball}{\mbox{\textsc{Crystalball}}}
\newcommand{\pybdsf}{\mbox{\textsc{PyBDSF}}}
\newcommand{\uJy}{\mu\mathrm{Jy}}

\newcommand{\revone}[1]{#1}

\title[RATT PARROT]{The RATT PARROT: serendipitous discovery of a peculiarly scintillating pulsar in MeerKAT imaging observations of the Great Saturn--Jupiter Conjunction of 2020. I. Dynamic imaging and data analysis}

\author[Smirnov, O.M. et al.]{O.~M.~Smirnov$^{1,2,3}$\thanks{E-mail: o.smirnov@ru.ac.za},
B.~W.~Stappers$^{4}$,
C.~Tasse$^{5,1}$,  
H.~L.~Bester$^{2,1}$, 
H.~Bignall$^6$\thanks{Visiting Scientist at CSIRO Space and Astronomy, 26 Dick Perry Ave, Kensington, WA 6151, Australia},
M.~A.~Walker$^6$, 
\newauthor
M.~Caleb$^{7,8}$, 
K.~M.~Rajwade$^{9}$,
S.~Buchner$^{2}$, 
P.~Woudt$^{10}$,
M.~Ivchenko$^{11}$,
L.~Roth$^{11}$,
\newauthor
J.~E.~Noordam$^{12,9}$\thanks{ASTRON (retired)}, 
F.~Camilo$^2$ 
\\
$^{1}$Centre for Radio Astronomy Techniques and Technologies (RATT), Department of Physics and Electronics, Rhodes University, Makhanda, 6140,\\South Africa\\
$^{2}$South African Radio Astronomy Observatory, Black River Park North, 2 Fir St, Cape Town, 7925, South Africa\\
$^{3}$Institute for Radioastronomy, National Institute of Astrophysics (INAF IRA), Via Gobetti 101, 40129 Bologna, Italy \\
$^{4}$Jodrell Bank Centre for Astrophysics, Department of Physics and Astronomy, The University of Manchester, Manchester, M13 9PL, UK\\
$^{5}$GEPI, Observatoire de Paris, CNRS, Université Paris Diderot, 5 place Jules Janssen, F-92190 Meudon, France\\
$^{6}$Manly Astrophysics, 15/41-42 East Esplanade, Manly, NSW 2095, Australia\\
$^{7}$Sydney Institute for Astronomy, School of Physics, The University of Sydney, NSW 2006, Australia\\
$^{8}$ASTRO3D: ARC Centre of Excellence for All-sky Astrophysics in 3D, ACT 2601, Australia\\
$^{9}$Netherlands Institute for Radio Astronomy (ASTRON), PO Box 2, Dwingeloo, 7990 PD, The Netherlands\\
$^{10}$Department of Astronomy, University of Cape Town, Private Bag X3, Rondebosch, 7701, South Africa\\
$^{11}$Space and Plasma Physics, KTH Royal Institute of Technology, SE-100 44, Stockholm, Sweden\\
$^{12}$Madroon Community Consultants, Pesse, 7933 RA, The Netherlands\\
}

\date{Accepted XXX. Received YYY; in original form ZZZ}

\pubyear{2023}

\begin{document}
\label{firstpage}
\pagerange{\pageref{firstpage}--\pageref{lastpage}}
\maketitle

\begin{abstract}
We report on a radiopolarimetric observation of the Saturn--Jupiter Great Conjunction of 2020 using the MeerKAT L-band system, 
\revone{initially carried out for science verification purposes, which yielded a serendipitous discovery of a pulsar.}
The radiation belts of Jupiter are very bright and time variable: coupled with the sensitivity of MeerKAT, this necessitated development of dynamic imaging techniques, reported on in this work. We present a deep radio ``movie'' revealing Jupiter's rotating magnetosphere, a radio detection of Callisto, and numerous background radio galaxies.
We also detect a bright radio transient in close vicinity to Saturn, lasting approximately 45 minutes. Follow-up deep imaging observations confirmed this as a faint compact variable radio source, and yielded detections of pulsed emission by the commensal MeerTRAP search engine, establishing the object's nature as a radio emitting neutron star, designated PSR\,J2009$-$2026. A further observation combining deep imaging with the PTUSE pulsar backend measured detailed dynamic spectra for the object. While qualitatively consistent with scintillation, the magnitude of the magnification events and the characteristic timescales are odd. We are tentatively designating this object a \emph{pulsar with anomalous refraction recurring on odd timescales} (PARROT).
As part of this investigation, we present a pipeline for detection of variable sources in imaging data, with dynamic spectra and lightcurves as the products, and compare dynamic spectra obtained from visibility data with those yielded by PTUSE. We discuss MeerKAT's capabilities and prospects for detecting more of such transients and variables.
\end{abstract}





\section{Introduction}

Jupiter has been an iconic object for astronomers from prehistoric times. Galileo's discovery of the Jovian moons (1610) was a key piece of data for the Copernican heliocentric model. With the discovery of radio emission from the planet by \citet{fb-jupiter}, and its modulation by interactions with Io \citep{bigg-jupiter-io}, Jupiter became a fascinating target for radio astronomers. The ``Great'' Conjunction of 2020 placed Saturn and Jupiter in their closest proximity since the time of Galileo. This provided a unique \revone{science verification opportunity for MeerKAT, with two radio-emitting planets in the same field of view. Although no new scientific insights were expected, the technical 
challenge of the observation provided a great opportunity to stress-test both the telescope systems and our software pipelines.} In L-band, Jupiter is an extremely bright radio source, largely owing to the so-called \emph{decametric} emission produced by cyclotron processes within its radiation belts \citep{dh-jupiter}. This emission is also strongly polarized, highly time-variable, and spatially resolved by MeerKAT. Combined with the proper motion of the planets relative to the background sources, this observation therefore represented a \emph{dynamic imaging} challenge not usually faced by conventional radio interferometric imaging.
\revone{Presented in movie form, the observations yielded a seredipitous discovery of a transient source in close vicinity to Saturn. Follow-up observations of this source confirmed its nature as a radio emitting neutron star.}


With the advent of large field-of-view SKA precursors such as MeerKAT, LOFAR and ASKAP, the search for transients has become a burgeoning industry. All three of these precursor telescopes are running large science projects that are successfully detecting new transients. Some examples from the ThunderKAT LSP of MeerKAT are reported by \citet{thk-andersson,andersson-citizen-science,thk-driessen1,thk-driessen2}, from the ASKAP VAST survey by \citet{askap-wang}, and from the LOFAR Transients KP by \citet{lofar-stewart}. Typically, such discoveries have been either triggered 
by dedicated transient search engines such as MeerTRAP, or targeted through imaging strategies optimized for transient detection, involving multi-epoch observations of the same field. Examples of these include \citet{thk-caleb} and \citet{nhw-transient,nhw-transient2}. Circular polarization, common in pulsars but rare elsewhere in the sky, can also be a powerful pulsar imaging signature \citep{kaplan-transient}. The detection reported here is unusual in having a pulsar announce itself in total intensity imaging so prominently.

MeerKAT's high instantaneous sensitivity and large FoV make it an ideal instrument for detecting transients and source variability on short timescales, using conventional synthesis imaging observations, so this discovery is perhaps not as serendipitous as it appears at first sight. One of the aims of this work is to illustrate the capability of the instrument for such transient searches. The final aim of this work is to present \revone{data reduction techniques developed for the} dynamic imaging observations of Jupiter.

This paper is structured as follows. Section~\ref{sec:obs} gives an overview of the observations. Section~\ref{sec:dynimg} discusses the ``dynamic imaging'' data reduction for the Great Conjunction observation, which led to the initial PARROT discovery. \revone{Section~\ref{sec:followup} discusses data reduction of the the follow-up observations, and the extraction of data products for transient and variability analysis.} Section~\ref{sec:timing} discusses the results from the pulsar timing engine. Section~\ref{sec:discuss} contains a preliminary discussion of the phenomenology of the PARROT. The latter seems highly irregular and meriting of a deeper investigation, which is deferred to a follow-up Paper~II. We end the paper with a summary and conclusions.



\section{Summary of observations}
\label{sec:obs}

\begin{table*}
    \begin{tabular}{c|c|c|c|c|c|c|c|c|l}
    label & UTC start & time, h & band & $N_d$ & SB ID & $\mu\mathrm{Jy}$ rms & $\mu\mathrm{Jy}$ peak & $\mu\mathrm{Jy}$ mean & commensal observations \\ 
    \hline
    L0 & 21-Dec-2020 08:18 & 9.4 & L   & 61 & 1608538564 & 20.0 / 7.0 & 5591$\pm127$ & 69 & none \\
    L1 & 31-May-2021 20:08 & 10.0 & L   & 62 & 1622491578 & 3.0 & 517$\pm148$ & 66 & MeerTRAP: no pulses \\
    L2 & 20-Jun-2021 19:14 & 10.0 & L   & 60 & 1624216341 & 2.9 & 750$\pm146$ & 86 & MeerTRAP: 5 pulses \\
    L3 & 27-Jul-2021 17:02 & 10.0 & L   & 29 & 1627405233 & 6.1 & 1250$\pm338$ & 57 &  none \\
    U0 & 7-Jul-2021	 02:07 & 1.3 & UHF & 60 & 1625623568 & 10.0 & 422$\pm153$ & 102 & MeerTRAP: no pulses \\
    U1 & 27-Jul-2021 17:02 & 10.0 & UHF & 29 & 1627405250 & 9.3 & 1145$\pm413$ & 138 & none \\
    U2 & 8-Aug-2021	 16:13 & 11.0 & UHF & 60 &  1628439081 & 4.9 & 1348$\pm161$ & 125 & MeerTRAP: 12 pulses \\
    \multirow{2}{*}{U3} & 4-Feb-2022 04:11 & 5.9 & UHF & 62 &  1643947704 & \multirow{2}{*}{4.9} &
                       \multirow{2}{*}{1922$\pm187$} & \multirow{2}{*}{231} & \multirow{2}{*}{continuous pulsar mode}    \\
       & 4-Feb-2022 10:19 & 4.6 & UHF & 61 & 1643969937 & & & & \\
    \end{tabular}
    \caption{\label{tab:obs}Summary of observations. $N_d$ is number of dishes available, SB ID is schedule block ID.
    Mean flux refers to the flux of the PARROT as measured in the full synthesis image; peak flux refers to peak flux in the lightcurve. L0 is the original ``detection observation'' of the Great Conjunction which was imaged dynamically per scan, here ``rms'' indicates the per-scan (15-min) sensitivity, and the overall mosaic sensitivity, the latter being limited by artefacts associated with Jupiter and the depth of per-scan deconvolution. All other observations were centred on the transient, and ``rms'' indicates the full synthesis sensitivity as measured at the centre of the field.
    Peak flux is measured from 8-second lightcurves. Note that the UHF observations are confusion-limited.
    L3 and U1 were carried out simultaneously by splitting MeerKAT into two subarrays. Data capture had to be restarted midway through U3, so it shows up as two schedule blocks which were concatenated and
    imaged together. Where the MeerTRAP backend was operating commensally, we indicate the pulse detections, if any.}
\end{table*}

\begin{figure}
    \includegraphics[width=.48\textwidth]{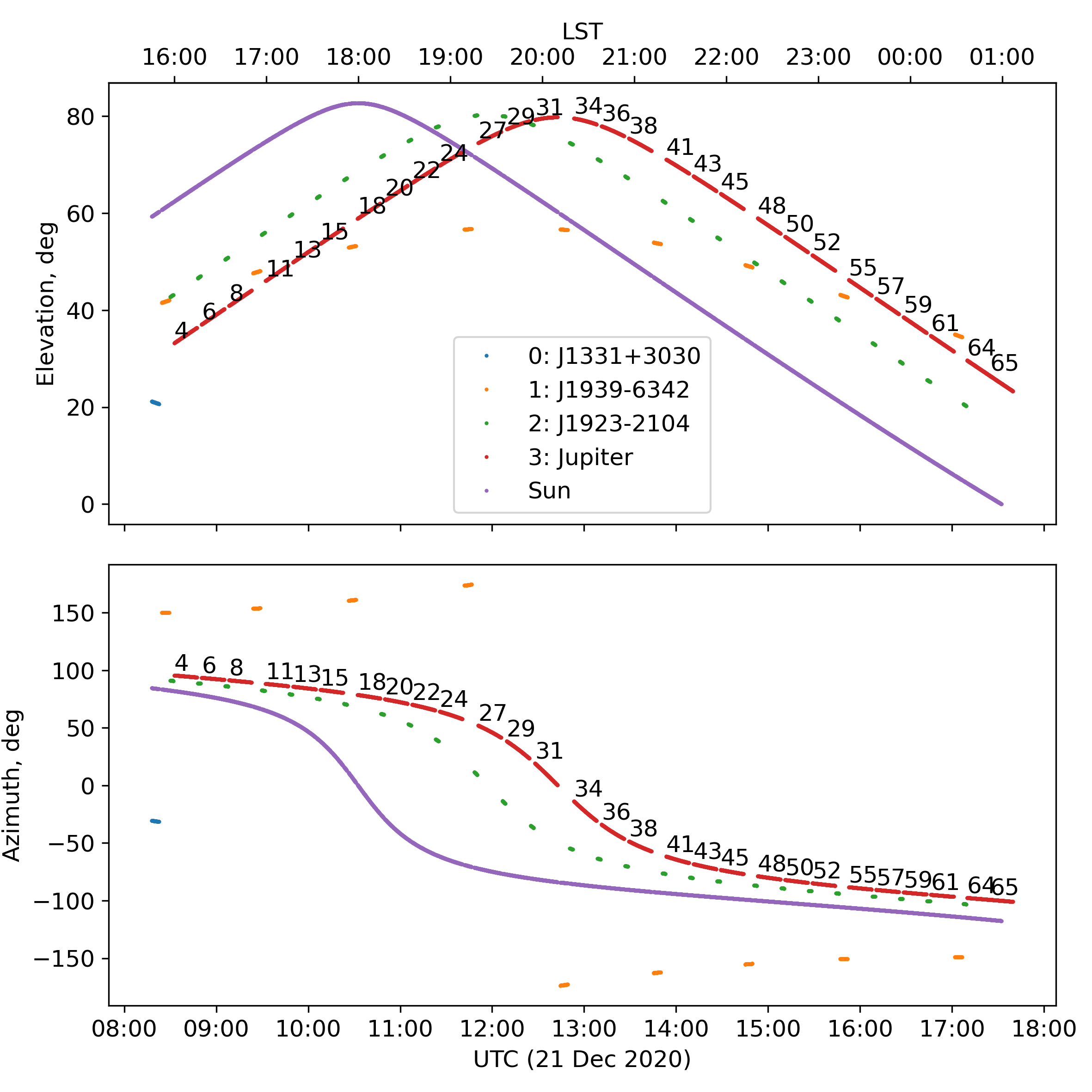}
    \caption{\label{fig:azel-tracks}Azimuth and elevation tracks for the Great Conjunction observation. Scans on Jupiter are plotted in red and labelled, scans on the gain calibrator are in green, the bandpass calibrator in orange. The position of the Sun is indicated in purple. 
    }
\end{figure}

The MeerKAT synthesis imaging observation of the Great Conjunction (schedule block ID 1608538564, proposal ID SSV-20200715-SA-01) commenced a few hours before the closest approach of Jupiter and Saturn, on 21 Dec 2020 08:18 UTC, using the L-band receiver (856--1712 MHz) in 4096 channel correlator mode, with 8 second integrations. The total duration of the observation was 9.4 hours, including calibrator scans. Due to the challenging and experimental nature of the imaging targets (i.e. two planets in the field of view moving relative to the background sources, Jupiter itself being very bright in L-band and variable in time), a conservative observing strategy was adopted, including 8 separate 15-minute visits to a bandpass calibrator (J1939-6342), and relatively short ``target'' scans of $14.8$ minutes each, interspersed with 2-minute scans on a nearby gain calibrator (J1923-2104). The observation started with a 5-minute scan of the polarization calibrator, 3C 286. During the target scans, both the antenna pointing and the correlator phase/delay centre tracked Jupiter. Figure~\ref{fig:azel-tracks} illustrates the resulting azimuth and elevation tracks.

\begin{figure*}
    \includegraphics[width=.9\textwidth]{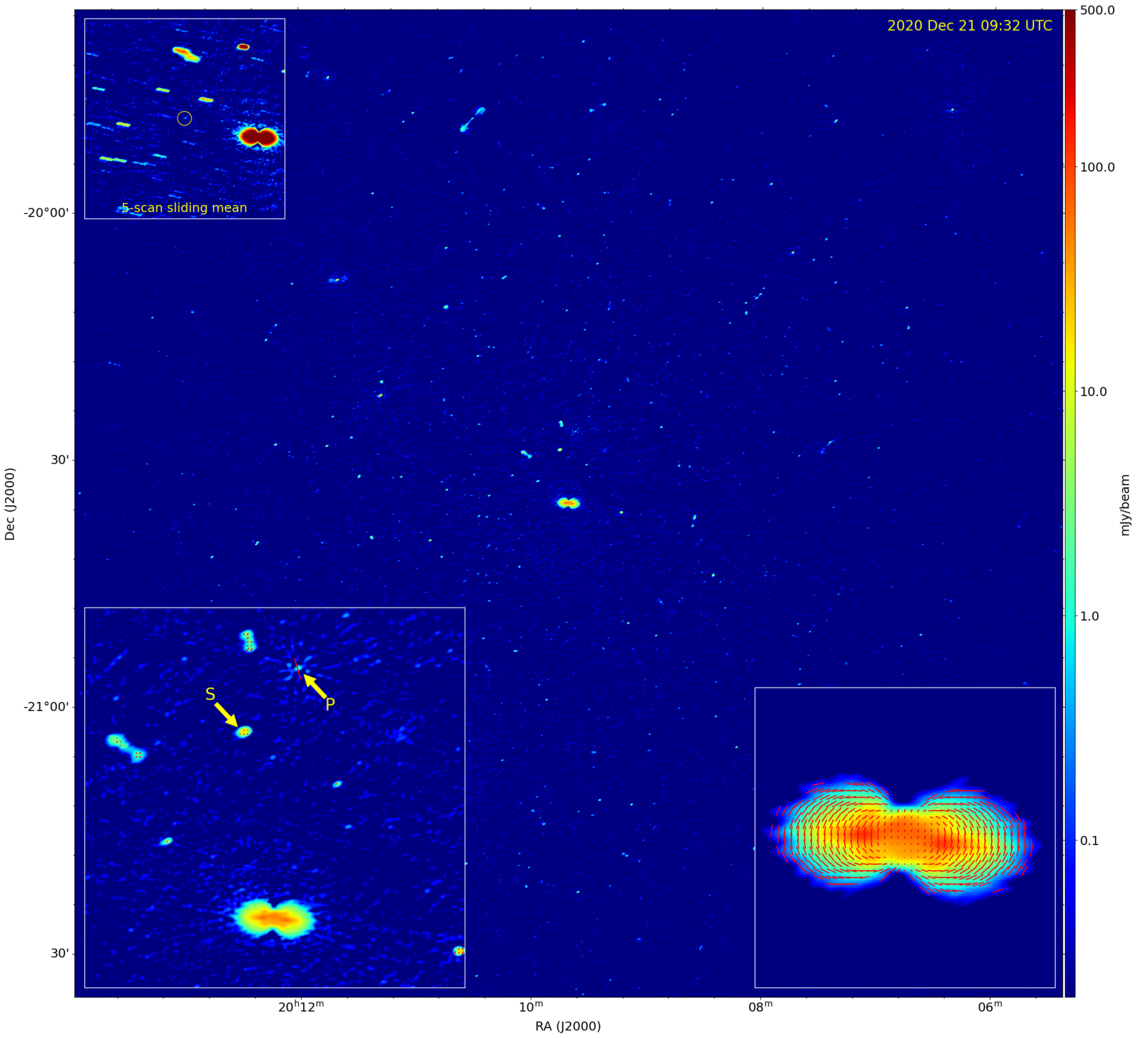}
    \caption{\label{fig:jove1-movie}
    {\bf (This is a placeholder for the animated figure, which can also be accessed via 
    \url{https://ratt.center/parrot}.)}
    The Great Conjunction observation, presented in per-scan movie form. Jupiter is at the centre. 
    Bottom left inset: zoom into a $6\arcmin \times 6\arcmin$ region containing Jupiter, Saturn and the PARROT.  
    Bottom right inset: a $1\farcm5\times1\farcm5$ zoom into Jupiter, with observed $B$-field vectors overplotted. 
    Top left inset: mean image of 5 consecutive scans
    (6, 8, 11, 13, 15), image size is $7\farcm2\times7\farcm2$. Callisto is marked by a circle.
    }
\end{figure*}

\begin{figure*}
    \includegraphics[width=.9\textwidth]{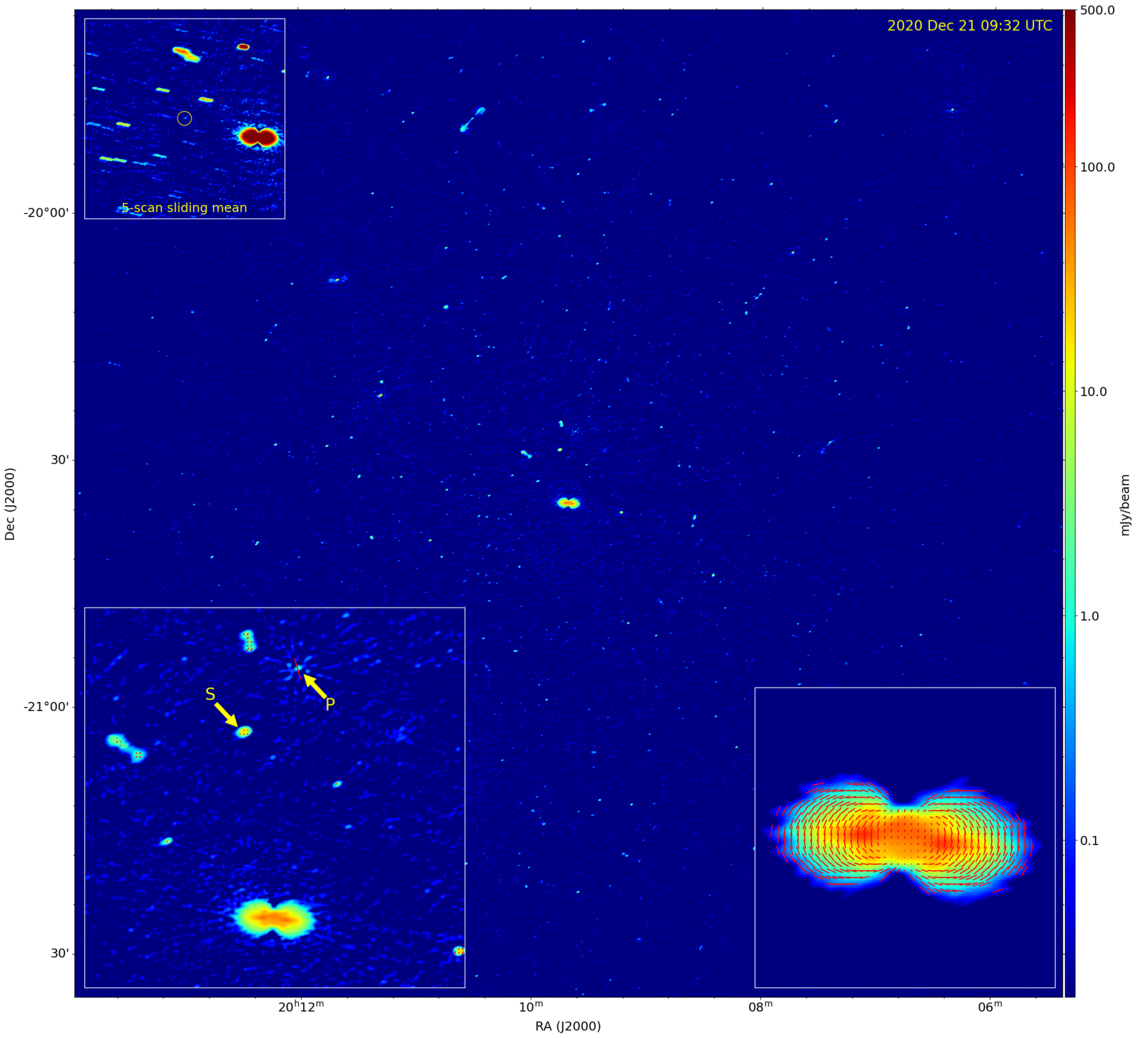}
    \caption{\label{fig:jove1}
    MeerKAT image of scan 11 of the Great Conjunction observation. Jupiter is at the centre. 
    The duration of the scan is 14.8 minutes. 
    Bottom left inset: zoom into a $6\arcmin \times 6\arcmin$ region containing Jupiter, Saturn (``S'') and the PARROT (``P'').  
    Bottom right inset: a $1\farcm5\times1\farcm5$ zoom into Jupiter, with observed $B$-field vectors overplotted. 
    Top left inset: mean image of 5 consecutive scans
    (6, 8, 11, 13, 15), image size is $7\farcm2\times7\farcm2$. Callisto is marked by a circle.
    }
\end{figure*}

\revone{The dynamic nature of the targets precluded normal static interferometric imaging where a single image of the field would be constructed. Instead, we created a ``movie'' composed of per-scan images (Fig.~\ref{fig:jove1-movie}). This showed that in radio, the Great Conjunction was far more eventful than expected. In addition to two distinctive FR2 \citep[Fanaroff-Riley Class II,][]{FR} radio galaxies joining the conjunction, we also observed a new radio transient in close vicinity. The transient manifested itself as a relatively bright (peaking at 5.6 mJy), almost 100\% polarized point source, which was detectable in three 15-minute snapshot images. Its discovery was thus a serendipitous by-product of doing dynamic imaging.}

The nature of the transient was difficult to establish from this initial observation and so several follow-up imaging observations were conducted (Table~\ref{tab:obs}) in both L-band and UHF (544--1088 MHz). These also employed 4096-channel mode and 8 second integrations (with the benefit of hindsight, given the pulsar nature of the transient, 2 second integrations would have been more useful, but this did not become obvious until later.) Since by this time Jupiter and Saturn had moved to different positions in the sky, these follow-up observations were conducted in a more conventional imaging mode, with the array tracking the fixed RA/Dec position of the suspected transient during the 30-minute target scans, interspersed with 2-minute scans of the gain calibrator (J2007-1016). J1939-6342 was used as the bandpass calibrator, with 3 or 4 scans of 15 minutes each, and 3C 286 as the polarization calibrator, with one 15-minute scan.

The \revone{object} was reliably detected in all follow-up observations as a variable point source with an average flux density of $58\sim86$ $\mu\mathrm{Jy}$ in L-band, and $102\sim231$ $\mu\mathrm{Jy}$ in UHF, with its lightcurves exhibiting occasional peaks \revone{(see Sect.~\ref{sec:followup})}. During observation \revone{U2}, the MeerTRAP transient search backend detected\footnote{\revone{After reprocessing, pulses were also detected in observation L2, but this was not automatically picked up at the time of the observation.}} several pulses in the coherent beam containing the \revone{object}. Since the times of these pulses coincided with peaks in the lightcurves, the pulsar-like nature of the \revone{object} became evident. A final observation (U3) was therefore performed commensally with the PTUSE pulsar backend operating in search mode. The results of this are discussed further in Sect.~\ref{sec:timing}.

\section{Imaging the Great Conjunction}
\label{sec:dynimg}

The Great Conjunction observation was highly non-standard in that the telescope continuously tracked a moving and variable object (namely, Jupiter) during the target scans -- that is, the pointing and phase direction changed with every integration. This type of observation is \revone{difficult to concisely express} in terms of the Measurement Set definition \citep[MSv2]{msv2}, since MSv2 specifies a {\tt FIELD} subtable containing fixed RA/Dec coordinates, and a conformant MS would thus require a separate field entry for every integration, while most tools expect a target to be a single field. The observatory adopted a workaround similar to that used by NRAO for planetary observations:\footnote{\url{https://casa.nrao.edu/docs/taskref/fixplanets-task.html}} the {\tt FIELD} subtable had a single entry specifying the position of Jupiter at the \emph{start}\/ of the observation, while the correlator phase centre (and computed $uvw$ coordinates) actually tracked the correct position of Jupiter throughout the observing run. Additional steps were then inserted into the processing workflow (see below) to accommodate this.

This complication did not affect the initial reference calibration process (a.k.a. 1GC), since the calibrator scans tracked fixed positions on the sky, as for any conventional imaging observation. We used the \caracal\ pipeline \citep{caracal} to do the reference calibration. This consisted of the following steps:\footnote{The full configuration file used with \caracal\ is available from the PARROT project page at \rraturl.}

\begin{itemize}
    \item Splitting out the calibrator scans;
    \item RFI flagging on the calibrator data using the \tricolour\footnote{\url{https://github.com/ratt-ru/tricolour}} package;
    \item Bandpass, fluxscale and gain calibration (via the standard CASA tasks);
    \item Polarization leakage calibration using J1939-6342, and crosshand phase calibration using 3C 286 (via the standard CASA tasks);
    \item Splitting out the target data and applying calibration solutions, averaging the target data down to 1024 channels;
    \item RFI flagging on the target data using \tricolour.
\end{itemize}

In a normal workflow, this would have been followed by full synthesis imaging and self-calibration of the target data. In this case it was 
\revone{obviously not practical} -- full synthesis imaging \revone{completely blurred Jupiter's radiation belts due to the planet's axial tilt}, \revone{smeared the image of Saturn due to its relative motion}, and produced long smeared tracks for all the background sources, which were drifting past as the telescope tracked Jupiter, at a rate of $0\farcs3/\mathrm{min}$.

The target data was therefore split up by scan, and each (14.8-minute) scan was treated as an independent imaging problem. The ``field'' direction corresponding to the scan was set to the position of Jupiter at the mid-point of the scan.
The sky drift over each scan was about $\pm 2\farcs2$, i.e. a \revone{substantial} fraction of the synthesized beam ($\sim 6\arcsec$). 
\revone{Nonetheless, this was short enough to eventually allow for artefact-free per-scan images (Fig.~\ref{fig:jove1}).}

\subsection{Per-scan polarised selfcal pipeline}
\label{sec:jove-pipeline}

The eventual pipeline consisted of the following steps:

\newlength{\plotheight}
\setlength{\plotheight}{.32\textwidth}

\begin{figure*}
    \begin{tabular}{@{}c@{ }c@{ }c@{}}
        \includegraphics[height=\plotheight,trim={0 0 92px 0 0},clip]{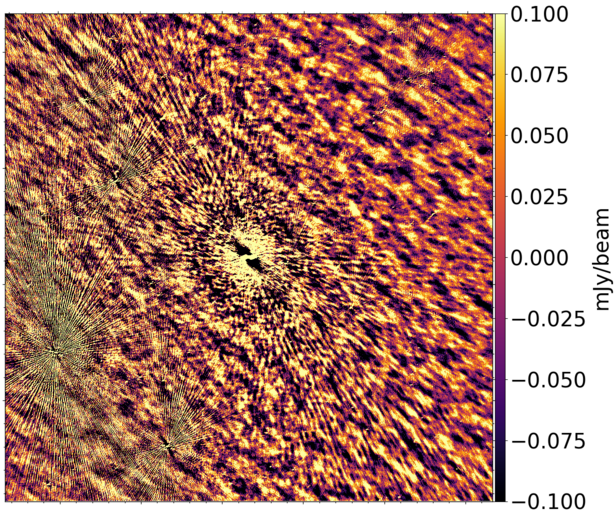} &
        \includegraphics[height=\plotheight,trim={0 0 92px 0 0},clip]{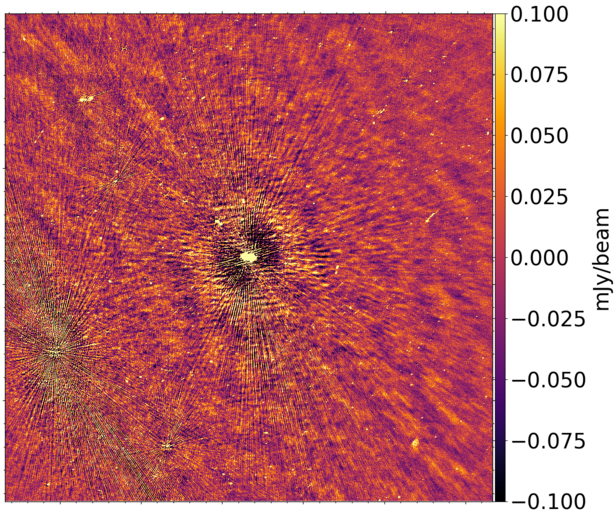} &
        \includegraphics[height=\plotheight]{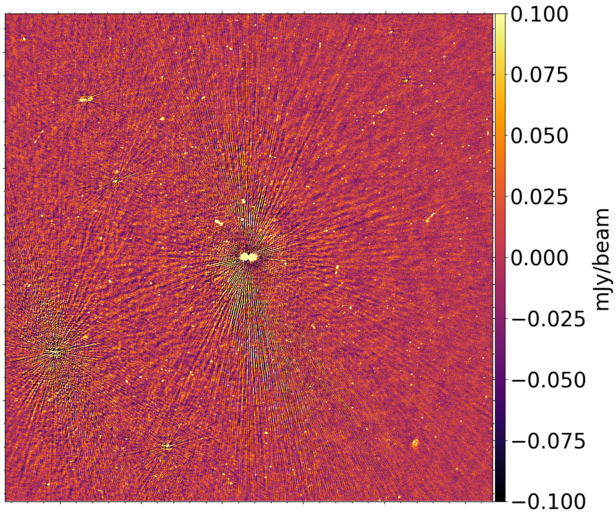} \\
        \includegraphics[height=\plotheight,trim={0 0 92px 0 0},clip]{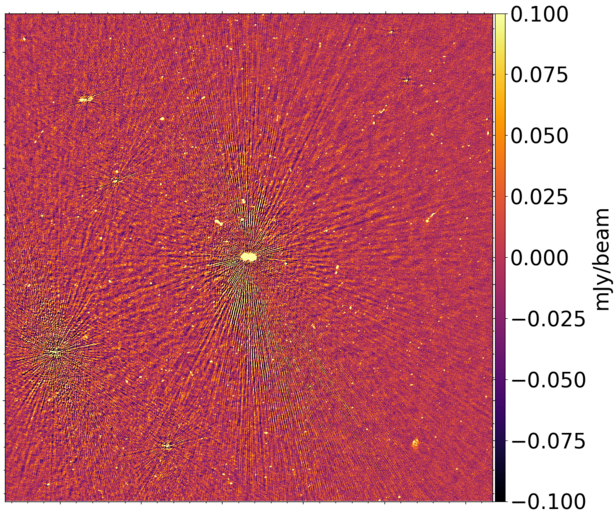} &
        \includegraphics[height=\plotheight,trim={0 0 92px 0 0},clip]{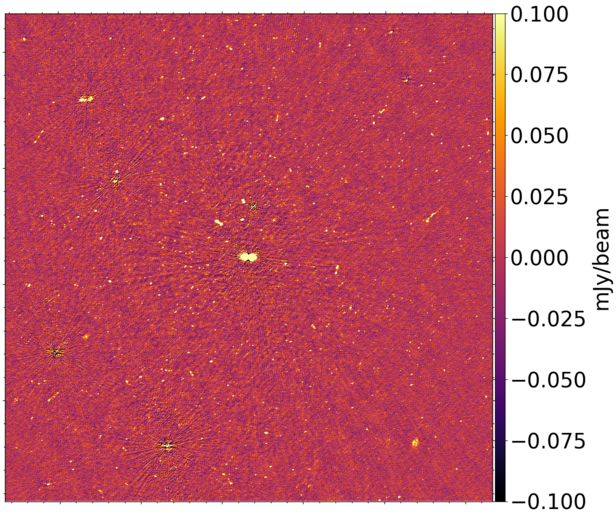} \\
    \end{tabular}
    \caption{\label{fig:scan11}The various stages of self-calibration for scan 11. Top left: initial image after 1GC. Top centre: 1 round of phase \& delay selfcal. Top right: 2 rounds of phase \& delay selfcal. Bottom left: 3 rounds of phase \& delay selfcal. Bottom centre: peeling and 4 rounds of phase \& delay selfcal, followed by 3.7-minute imaging. Colour scale is locked across all images. Images are $1\fdg4 \times 1\fdg4$.
    }
\end{figure*}

\noindent{\bf Initial imaging} in Stokes $I$ was done using the \wsclean\ imager, with a robustness value of 0, using 8 imaging subbands with joint-channel cleaning and 4th-order spectral polynomial fitting. Only single-scale CLEAN was used\footnote{Multiscale cleaning did not produce any appreciable improvement in the images at any stage, and was a lot slower, so single-scale was used throughout the pipeline.}. The resulting images (see e.g. Fig.~\ref{fig:scan11}, top left) were relatively poor, but proved sufficient to start the self-calibration process. The images were also used to construct a cleaning mask for the subsequent step, using the \breizorro\footnote{\url{https://github.com/ratt-ru/breizorro}} masking tool with a threshold of $10\sigma$.

\noindent{\bf Three rounds of phase \& delay selfcal} in full Stokes $IQUV$. Both a per-antenna phase offset and a delay term (with a solution interval of 8 seconds, i.e. a single integration, Jupiter being bright enough to provide sufficient SNR for this) were solved for 
at each round, using the \cubical\footnote{\url{https://github.com/ratt-ru/CubiCal}} calibration suite \citep{cubical}. Stokes $IQUV$ images were generated with \wsclean\ at each round, and fed into \breizorro\ (using a threshold of $6.5\sigma$) to produce a mask for the next round of cleaning. The resulting images from each round, for one scan, are shown in Fig.~\ref{fig:scan11} top centre, top right, and bottom left. 

The self-calibration process yielded a number of interesting insights. Firstly, three rounds of selfcal were really necessary for many of the scans to converge to a \revone{relatively artefact-free image} (though perhaps not for the scan shown in Fig.~\ref{fig:scan11}). Secondly, when we restricted ourselves to Stokes $I$ imaging in an early version of the pipeline, \revone{self-calibration of some scans failed to converge. We established that} this was due to Jupiter's high degree of linear polarization -- the unmodelled Stokes $Q$ signal (in the parallel-hand correlations) was interacting with the independent $X/Y$ delay solutions, resulting in extended imaging artefacts around Jupiter. 
Our eventual solution was to add a so-called \emph{unislope} solver to \cubical, which solves for independent $X/Y$ phase offsets, but a single common phase slope (i.e. delay) term. This, in combination with full Stokes imaging, allowed selfcal to converge in all scans (strictly speaking, only Stokes $IQ$ imaging is necessary for the selfcal, since the $UV$ signal does not enter the $XX$ and $YY$ correlations, but our final imaging product required full Stokes anyway).   

\noindent{\bf Peeling and selfcal}, i.e. direction-dependent calibration using differential gains \citep{rime3}, 
was then required to deal with the brightest off-axis source in the field (see Fig.~\ref{fig:scan11}, eight o'clock position), which was contributing radial artefacts to the map. Peeling was performed using \cubical, using the following form of the radio interferometer measurement equation \citep[RIME:][]{rime1}:

\begin{equation}
    \label{eq:peel1}
    V_{pq} = K_p(M_{pq}-M^{(1)}_{pq} + dE_{p}M^{(1)}_{pq}dE_{q}^H)K_q^H
\end{equation}

Here, $V_{pq}$ is the observed $2\times2$ visibility matrix data for the baseline formed by antennas $p$ and $q$; $M_{pq}$ is the visibility model for the field (as predicted by \wsclean\ from the last round of selfcal); $M^{(1)}_{pq}$ is a visibility model for the offending source; $K_{p}$ is a solvable phase-and-delay Jones term for antenna $p$, and $dE_p$ is a solvable differential gain Jones term (diagonal complex). The solution intervals for the $dE$ term were 16 timeslots and 128 frequency channels. The visibility model $M^{(1)}_{pq}$ was obtained as follows: we ran \wsclean\ in Stokes $I$-only mode and generated a source list containing the clean components and their spectral polynomial coefficients (note that the \wsclean\ source list feature is not available in full-Stokes mode), then selected the components within a small circular region around the offending source, then converted them into visibilities via a direct Fourier transform (DFT) using the \crystalball\footnote{\url{https://github.com/caracal-pipeline/crystalball}} tool \citep{crystalball}. The corrected visibilities are then formed up as

\begin{equation}
    \label{eq:peel-correct}
    V^\mathrm{(corr)}_{pq} = \tilde{K}^{-1}_p(V_{pq} - \tilde{K}_p \tilde{dE}_{p}M^{(1)}_{pq}\tilde{dE}_{q}^H \tilde{K}_q^H)\tilde{K}^{-H}_q,
\end{equation}

where $\tilde{K}_p$ and $\tilde{dE}_p$ are the calibration solutions derived by \cubical.

\noindent{\bf Subinterval imaging.} Post-peeling, the only prominently remaining artefacts in the maps were radial spokes centred on Jupiter. After some experimentation, we discovered that these could be suppressed by splitting each scan into four subintervals (of 3.7 min duration each), and imaging (in full Stokes) and deconvolving each subinterval independently (\wsclean\ provides an {\tt -intervals-out} parameter that makes this straightforward). The (inverse variance weighted) mean image over the four subintervals is given by Fig.~\ref{fig:scan11}, bottom centre. We conclude that the main reason for the artefacts in the full-scan (14.8 min) images is rapid (minute-timescale) variability in Jupiter's radiation belts, which per-subinterval deconvolution is able to account for.

\noindent{\bf Residual flagging.} The residual visibilities from the previous step (formed up by subtracting the model column generated during subinterval imaging from the corrected data column) were passed to \tricolour\ for another round of flagging. This picked up some additional faint RFI in several scans.

\noindent{\bf Repeeling.} The peeling+selfcal step was repeated, using the visibility model generated by the previous imaging step. This model (presumably) accounts for rapid variations in the radiation belts, but does not include the previously peeled source. The RIME for this step is, therefore, slightly different:

\begin{equation}
    \label{eq:peel2}
    V_{pq} = K_p(M_{pq} + dE_{p}M^{(1)}_{pq}dE_{q}^H)K_q^H.
\end{equation}

This resulted in a marginal improvement in the maps for some of the scans. In retrospect, it is questionable whether this final improvement is worth the extra processing time.

\noindent{\bf Final imaging.} The subinterval imaging step is repeated once more, using the corrected visibilities from the repeeling step. The inverse variance weighted mean Stokes $IQUV$ images are taken to be the final imaging product for that scan. We also generate residual visibilities once more, since these are needed to generate the dynamic spectra for the PARROT (Sect.~\ref{sec:dynspectra}) later. Note that in the final version of this workflow (once the PARROT had been discovered), the PARROT was removed from the cleaning mask at this stage, since we wanted its full signal present in the dynamic spectra.

\subsection{Pipeline automation}

The Great Conjunction observation consisted of 28 target scans. It would have been impractical to rerun the procedure described above for each scan by hand, yet on the other hand, it is sufficiently non-standard to preclude the use of any existing calibration and imaging pipeline. We essentially needed to build a custom pipeline from scratch. For a single one-off observation, this would have been quite the onerous task, especially considering the number of diverse software packages that needed to be made to interact together.

Fortuitously, the processing of this observation coincided with a complete rewrite of the \oldstimela\ pipeline framework \citep{stimela}. The process drew on three years of experience of the RATT group with \oldstimela\ (among other things, it is the underlying technology for the \caracal\ pipeline), 
as well as on the new challenges posed by the Great Conjunction data reduction. The result, \stimela\footnote{\url{https://github.com/caracal-pipeline/stimela2}} (Smirnov et al., in prep), is proving to be an extremely flexible \emph{workflow management} framework. It allows complex workflows, such as the one described above, to be expressed in terms of human-readable and succinct YaML recipes, and allows for a high degree of \emph{modularity} (i.e. support for libraries of recipes and task definitions), \emph{composability} (recipes may be included as sub-steps of other recipes, YaML configurations can be included in one another), and \emph{reproducibility} (recipes can be shared and rerun on other compute environments, while the use of container technology can ensure that the exact same versions of all relevant software is used). 

The \stimela\ recipes for the pipeline described in Sect.~\ref{sec:jove-pipeline} can be found on the PARROT project website. The technically inclined reader is urged to look at the YaML files, as the sequence of steps defined therein matches the procedure described above fairly closely, and supplies the finer details omitted in this manuscript. 

The two top-level recipes are the single-scan recipe, which can be run as e.g. 

\begin{verbatim}
$ stimela run jove-pol.yml scan=11 
\end{verbatim}

\noindent ...and the multi-scan recipe (which illustrates \stimela's for-loop feature):

\begin{verbatim}
$ stimela run jove-pol-loop.yml scan-list=[4,6,8,11]
\end{verbatim}

We leave further details to the \stimela\ paper (Smirnov et al., in prep) and technical documentation. Here we only wish to stress that the provided recipes
form an integral part of this work, and serve both to document the finer details of the calibration and imaging procedure, and to support a fully reproducible data reduction.

\subsection{Making movies: Callisto and the PARROT}

The output of the per-scan pipeline consists of $4\times28$ FITS images (four Stokes parameters and 28 scans). We have combined these images into an animation showing total intensity, fractional polarization and $B$-field vectors, as a function of time.
The full animation is presented\footnote{See also \url{https://ratt.center/parrot}} in Fig.~\ref{fig:jove1-movie}, and a single frame (corresponding to scan 11) is shown in Fig.~\ref{fig:jove1}. 

The animation illustrates quite a few fascinating aspects of this observation:

\begin{itemize}
    \item There is a very clear transient source lasting for approximately three scans (45 minutes) just above and to the right of Saturn (also indicated by an arrow in Fig.~\ref{fig:jove1}.) Follow-up observations, detailed below, have confirmed sporadic pulsed emission from the source, which suggests that this is a pulsar-type object. To our knowledge, this is the first direct discovery of a pulsar via single-epoch imaging. The discovery is the epitome of serendipity: in a normal multiple-hour synthesis, such a short duration transient would manifest itself as a mere faint unresolved source: it is only because we made an animation (and the object popped up in such an attention-grabbing time and place) that it became detectable.
    
    \item At 9.1 hours in duration, the observation covers almost a full Jovian day (9.9 hours). We can clearly see the effects of Jupiter's rotating magnetic field: the magnetic axis is tilted $\sim10\degr$ with respect to the rotation axis \citep{jupiter-magnetic-field}, which produces a corresponding tilt-and-rotation in the radiation belts. (It is somewhat fascinating that the Great Conjunction has also brought into conjunction two objects where a rotating magnetic field drives the radio emission, and where substantial circular polarization is present.) This behaviour is consistent with previous detailed observations of the radiation belts \citep[e.g.][]{Sault-jove}.
    
    \item The observed $B$-field vectors show an overall dipole around Jupiter. The two FR2's arriving on the scene are weakly polarized within their lobes. The PARROT shows an extremely high degree of both linear and circular polarization. These properties are to be expected for the respective classes of objects, and serve to confirm that our polarization calibration (and pol-selfcal) is yielding sensible results.

    \item Saturn appears rather less spectacular than its bigger cousin (or, indeed, than its own familiar optical manifestation). Its angular extent of $15\arcsec$ is only three resolution elements across. We do not detect (or resolve) the rings, which is not unexpected, given our resolution and the rings' low surface brightness at L-band \citep{zhang-saturn, depater-saturn}. We detect some linear polarization across the planetary disk, on the order of a few percent. We do not analyze Saturn further in this work.
    
    \item As a bonus detection, we observe a Galilean moon in radio. Callisto happened to be at a relatively large separation from Jupiter during the observation. It proved too faint to be detected in a single-scan image, but by taking a sliding mean (sliding window in time) of every 5 scans, we can detect a moving compact source at exactly the expected position (highlighted by an circle in the top left inset). Given that Callisto itself is smeared by its motion relative to Jupiter, an accurate flux measurement is non-trivial. A simple measure of peak pixel values gives us a range of $50\sim80$ $\uJy$.  
    
    Callisto's radio emission is expected to be purely thermal in origin. We are not aware of any direct measurements of its flux density at L-band, but extrapolation from \citet{callisto-butler} and \citet{callisto-camarca} suggests a brightness temperature of $\sim110$ K. The distance to Jupiter at the time of observation ($890\cdot10^6$ km) and the diameter of Callisto (4820 km) gives us a source extent 
    of $1\farcs1$. Combining that with the Rayleigh-Jeans approximation, we would predict Callisto to have a flux density of $\sim120  \uJy$ at our effective frequency of 1.28 GHz, which is somewhat discrepant with our peak pixel measurements. However, the latter are bound to be an underestimate due to Callisto's apparent motion over each scan. Since this is completely tangential to the present work, we defer a more accurate measurement as a bookmark for future follow-up.

    We do not directly detect the other three Galilean moons in these images. Given the increased level of residual deconvolution artefacts 
    closer to Jupiter, this is not surprising. The four moons have comparable brightness temperatures \citep{callisto-butler}. Callisto was the moon with the largest separation at the time. Europa was the second furthest out, but given its smaller size (3120 km), its expected flux density is less than half that of Callisto. Ganymede's flux density should be comparable to Callisto's, but it was much closer to Jupiter
    at the time. Finally, Io is somewhat smaller and fainter, and being the innermost moon, is the most challenging to detect in such observations. 
    
\end{itemize}

\begin{figure*}
    \includegraphics[width=.9\textwidth]{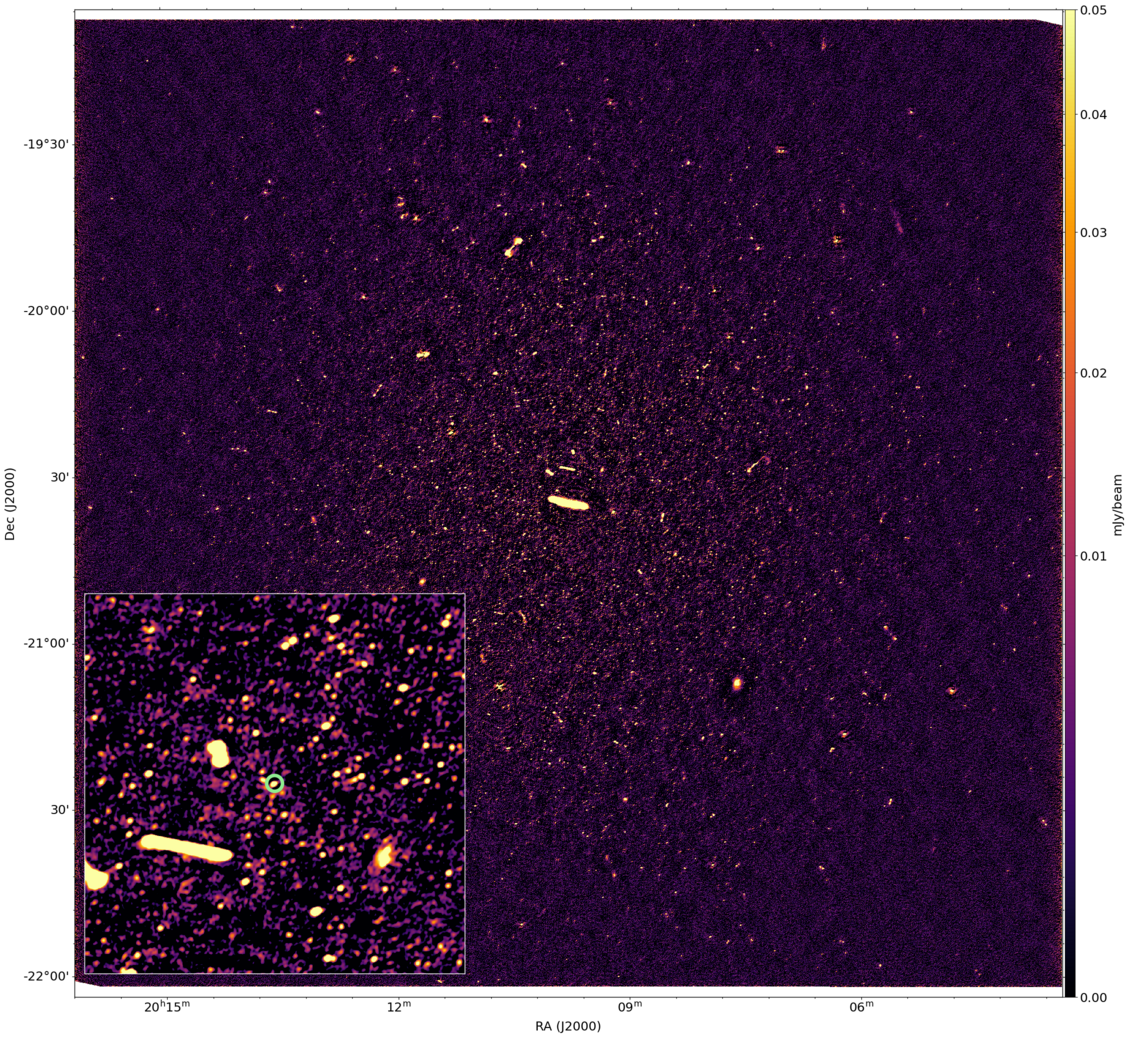}
    \caption{\label{fig:mosaic}
    MeerKAT mosaic of the Great Conjunction observation. Per-scan images have been combined into an image-plane mosaic.
    Jupiter and Saturn are smeared by their motion throughout the observation. The PARROT is circled.
    }
\end{figure*}

As an alternative way to present this information, we also construct an image-plane mosaic of the per-scan maps (Fig.~\ref{fig:mosaic}). This becomes a deep map (7 $\uJy$ sensititivity) of the extragalactic sources in the field, with Jupiter and Saturn smeared out by their motion. Note that the PARROT is clearly detected as a point source in this map.

\section{Imaging the follow-up observations}
\label{sec:followup}

Since the follow-up observations were conducted in a more conventional imaging mode, and did not feature an exotic source such as Jupiter, the initial data reduction procedures were correspondingly 
more straightforward. 1GC was performed using \caracal, following essentially the same procedure as described in Sect.~\ref{sec:dynimg}. 

The 1GC-corrected target data were then split out into a separate MS and fed through a custom selfcal pipeline implemented with \stimela\ (recipes available on project website). Details of this are described below. Finally, lightcurves of all sources in the field, and dynamic spectra of the PARROT were extracted.

\subsection{Imaging considerations for variability and transient studies}

The ultimate goal of the pipeline is to extract lightcurves and dynamic spectra for the PARROT (as well as, ultimately, other field sources, as the scope of our analysis widened.) Broadly speaking, we want to construct the deepest possible (time-averaged) sky model for the field, then subtract it from the visibilities so that we can analyze temporal and spectral variability at each target source position, while minimizing contamination by sidelobes from other sources. (This is not dissimilar to the continuum subtraction problem in spectral line work, with an extra time dimension thrown in.) The reduction strategy we eventually converged on is therefore somewhat different from conventional deep wideband continuum imaging. 

Firstly, we want to ensure spectral smoothness in the (subtracted) sky models, to avoid introducing unwanted spectral signatures into the dynamic spectra. Conventional wideband imaging is done over a few (and thus relatively wide) subbands, resulting in a stepwise spectrum for each source in the constructed sky model. The resulting errors tend to average down during regular imaging, but can impart unwanted signatures into the dynamic spectra of interest during sky model subtraction. 

Secondly, we want to be very conservative in our treatment of direction-dependent effects (DDEs). On the one hand, untreated DDEs contribute imaging artefacts, which translate to residual signal (with a temporal and spectral signature) when the sky model is subtracted. The best deep continuum images tend to employ DD-calibration in multiple directions to suppress the imaging artefacts. On the other hand, excessive degrees of freedom in DD solutions are known to cause flux suppression of unmodelled sources \citep{sob-harmful}, and given that the sky model is constructed from a deep multiple-hour synthesis image, any temporal variability becomes unmodelled by definition. Thus, the same process that yields the best continuum image will artificially suppress variability in the sources of interest! There is obviously a balance to be struck here.

Consider that the spectral signature of DDE-induced artefacts is modulated by the PSF sidelobes, which become increasingly random at increasing distance from the corrupted source. The contribution to the dynamic spectrum at any given target position from an ensemble of corrupted sources (that are imperfectly subtracted due to DDEs) is thus increasingly close to an additional Gaussian noise term, as long as there is no dominant DDE-contributing source nearby. In other words, we can tolerate DDEs from a collection of weaker sources -- they will, effectively, just increase the noise level -- but we do want to deal with the dominant DDE contributors. Fortunately, there is only one such dominant off-axis source in the field -- the same one that needed to be dealt with in the Great Conjunction images (Sect.~\ref{sec:dynimg}).

Thirdly, and for similar reasons, we want to be careful that our sky model does not pick up the remaining DDE-induced artefacts. Unmasked blind CLEANing \revone{is prone to picking up sidelobe peaks, and absorbing} some of their flux into the sky model. This can be mitigated via careful CLEAN masking.

Fourthly, since some of the observations were conducted in daytime (with U3 particularly close to the Sun), solar RFI was an issue. We dealt with this by imposing a $uv$-cut and an inner Tukey taper during imaging, on a per-observation basis. The settings for these were tweaked by examining the $uv$-amplitude maps produced at stage 4 of the pipeline (see below).

\subsection{Pipeline structure}
\label{sec:followup-pipeline}

This pipeline consists of the following steps:

\vspace{1ex}
\noindent{\bf Step 1. Initial imaging} is done using \wsclean. This is done with a robustness value of 0, using 8 imaging subbands for L-band observations (6 for UHF) with joint-channel cleaning and 4th-order spectral polynomial fitting, down to an auto-threshold of $2\sigma$. Only single-scale CLEAN was used. Initial versions of the pipeline employed \wsclean's auto-masking feature, but this tended to absorb sidelobe artefacts into the sky model (see discussion above). After some experimentation, we settled on using static \breizorro\ masks for the final version of the pipeline. The initial CLEANing mask for this step was constructed with \breizorro\ using a post-selfcal image produced by a previous pipeline run. The initial image is then flipped into visibilities to bootstrap the selfcal.

\vspace{1ex}
\noindent{\bf Step 2. Delay+phase selfcal} is done using \quartical\footnote{\url{https://github.com/ratt-ru/QuartiCal}}, the successor package to \cubical.
We used a solution interval of 1 timeslot for the delays and phase offsets. The resulting corrected visibilities are then imaged with \wsclean\ down to a threshold of $1\sigma$, using a \breizorro\ mask constructed from the initial image.

\vspace{1ex}
\noindent{\bf Step 2.1. Sky model smoothing.} The 8-band (6-band) sky model derived from the previous step was then upsampled into 256 channels using the \smops\footnote{\url{https://github.com/ratt-ru/smops}} tool, which takes in a set of per-band model images produced by \wsclean, and outputs a set of model images at a higher frequency resolution, using polynomial interpolation in frequency at each model pixel. The resulting models are then fed back into \wsclean\ to predict model visibilities at the increased spectral resolution. The value of 256 is a trade-off between smoothness and compute and disk space cost.

\vspace{1ex}
\noindent{\bf Step 2.2. A direction-dependent sky model} is constructed for the dominant off-axis source by running \crystalball\ to select and predict relevant clean components from the previous imaging run, exactly as described in Sect.~\ref{sec:dynimg}.

\begin{figure*}
    \includegraphics[width=.66\textwidth]{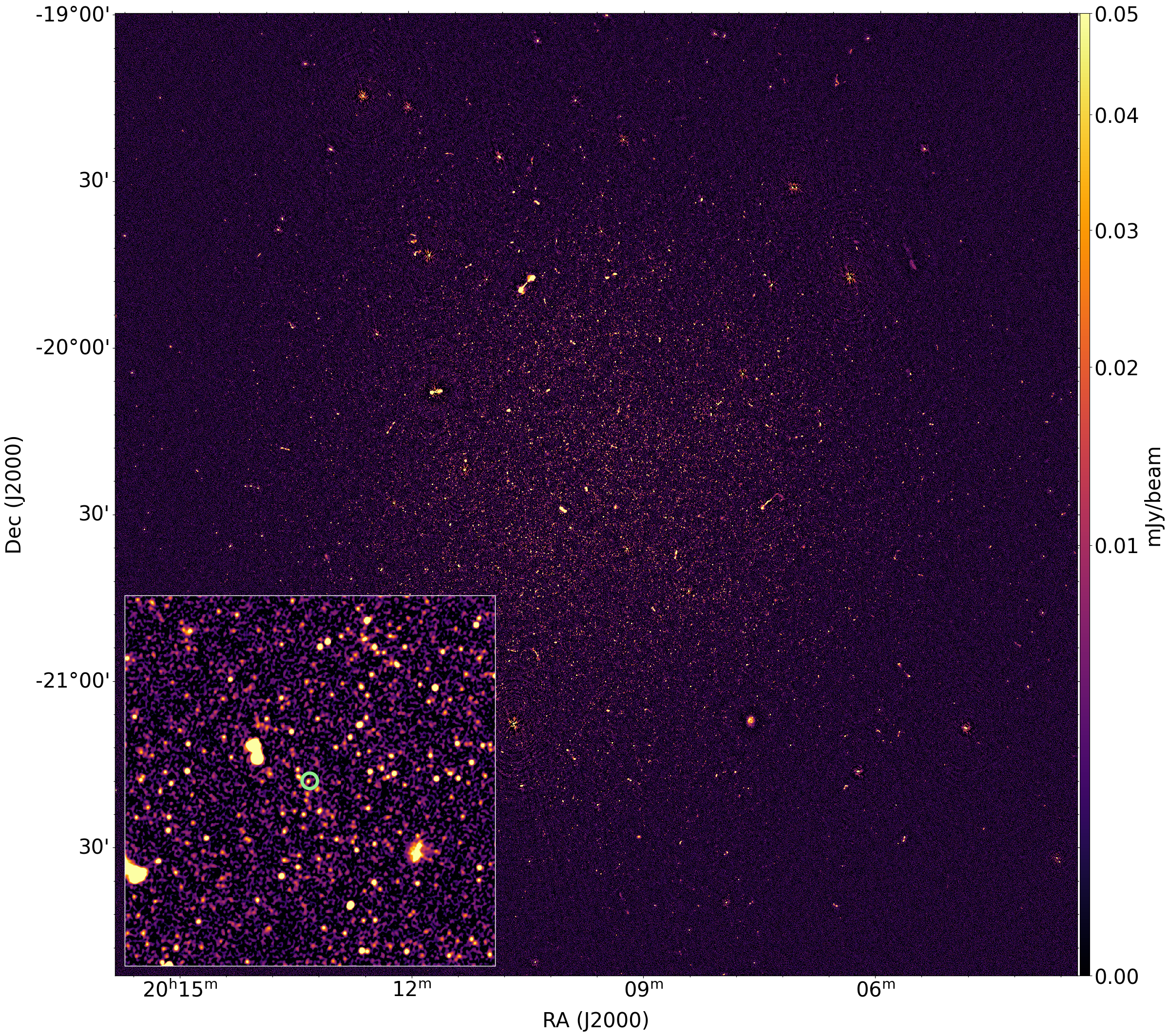}
    \caption{\label{fig:l1}MeerKAT image of a deep follow-up observation in L-band (observation L1). Inset shows zoom into the centre
    of the field. The PARROT is circled. 
    }
\end{figure*}

\begin{figure*}
    \includegraphics[width=.66\textwidth]{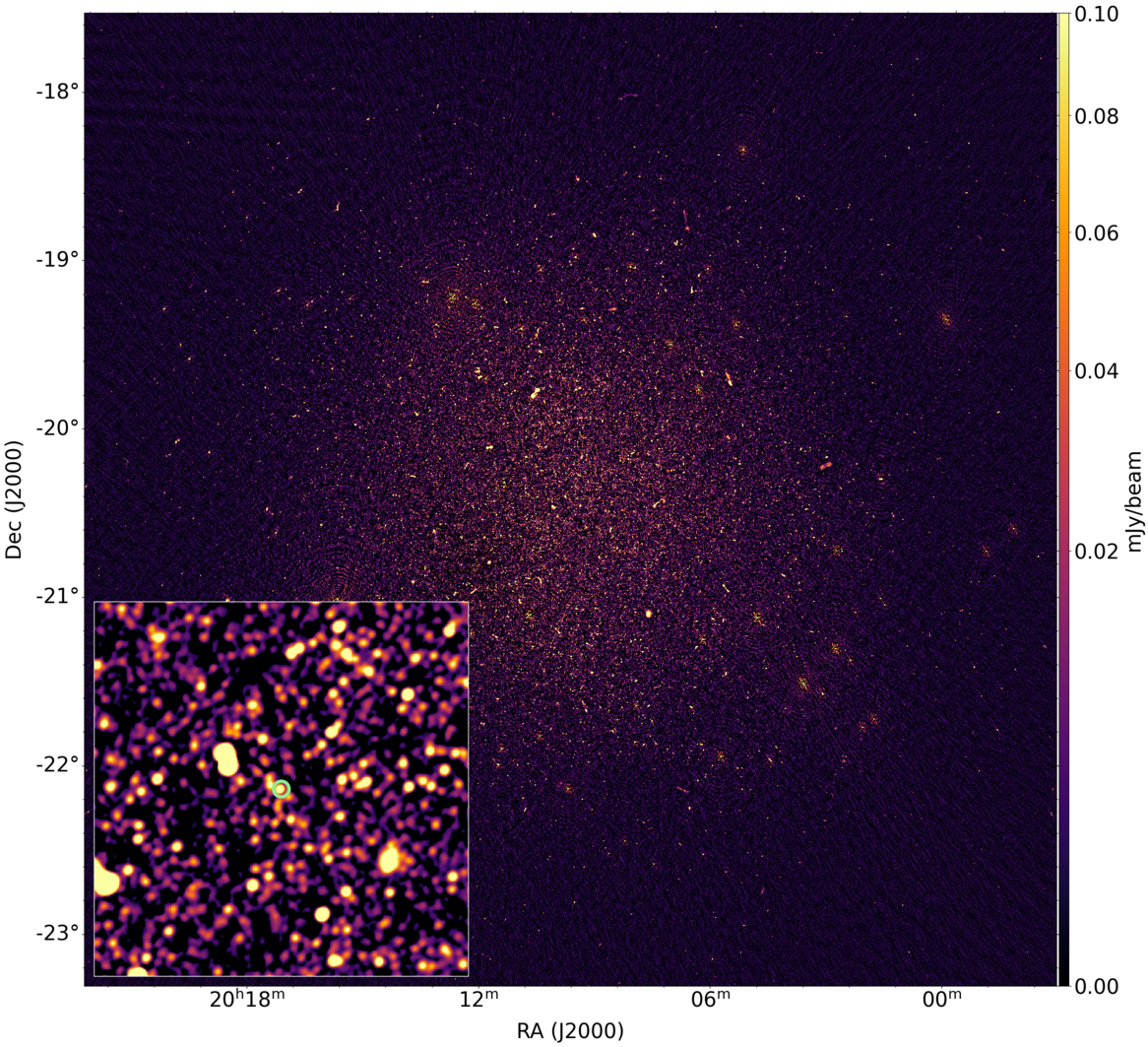}
    \caption{\label{fig:u3}MeerKAT image of a deep follow-up observation in UHF (observation U3). Inset shows zoom into the centre
    of the field. The PARROT is circled. 
    }
\end{figure*}

\vspace{1ex}
\noindent{\bf Step 3. Peeling \& selfcal} is then done via \quartical, using the spectrally smooth sky model and the DD sky model derived above, with the RIME given by eq.~\ref{eq:peel1}. \quartical\ produces corrected visibilities with the off-axis source peeled away, using eq.~\ref{eq:peel-correct}.
These are then fed into \wsclean\ to produce a deep image (Figs.~\ref{fig:l1}, \ref{fig:u3}). The sky model smoothing step is repeated, and an updated sky model is predicted.

\vspace{1ex}
\noindent{\bf Step 4. Peeling \& selfcal, second round} is done using the updated sky model. Since this no longer contains the off-axis source,
the version of the RIME employed is that of eq.~\ref{eq:peel2}. For this round, we tell \quartical\ to produce corrected residuals -- that is,
both the smooth sky model and the DD sky model are subtracted from the output visibilities. These are then reimaged with \wsclean\ in order to deconvolve any remaining faint sources not included in the subtracted sky models.

We then Fourier transform the \emph{residual} image from this last deconvolution round to obtain a map of the residual $uv$-amplitudes.
This is a valuable diagnostic product, since any faint RFI and bad individual baselines show up as bright peaks in the $uv$-plane. These maps were used to tweak the flagging and imaging settings (see discussion on solar RFI above) before re-running the pipeline. 

\vspace{1ex}
\noindent{\bf Step 5. Finalizing the ultimate residuals.} The model images from step 4 (containing the remaining faint sources) are converted back to visibilities using \wsclean, then subtracted from the corrected residual data generated by step 4. The result is a set of visibilities with our best possible integrated model of the sky subtracted. We call these the \emph{ultimate residuals}, and use them to construct the lightcurves and dynamic spectra below. These ultimate residuals contain three components:

\begin{itemize}
    \item Our science signal: the temporal and spectral variation of each source with respect to its subtracted model;
    \item Contaminating signal due to any remaining DDEs, and fainter sources below the deconvolution threshold;
    \item Thermal noise. 
\end{itemize}

We can then proceed to extract light curves and dynamic spectra for each source of interest, using the ultimate residuals. Note that a time-averaged and spectrally smooth model for each model source has been subtracted at this point, so the resulting lightcurves and spectra are ``mean-subtracted'' -- that is, they should show only variation around a mean of 0. The subtracted mean signal is contained in the sky models generated by the previous stages, and can be straightforwardly added back to the lightcurves afterwards, if needed.

\subsection{Variability measurements}

Our variability analysis was initially tightly focused on the PARROT. However, the need to discriminate between intrinsic variations, scintillation and instrumental effects widened the scope of our investigation to include other sources in the field.

The two primary data products for analysing variabilty are dynamic spectra and lightcurves. A dynamic spectrum, $S(t,\nu)$, represents the variation in the flux of a source as a function of time and frequency. A lightcurve, $S(t)$, is just the 1D version of this, representing the temporal variation in flux integrated over a frequency band. 

Given the position of a source and a set of ultimate residual visibilities, we can compute the (mean-subtracted) dynamic spectrum $S(t,\nu)$ by, essentially, beamforming: phase-shifting all the visibilities to the source position, then computing the (sensitivity-weighted) mean signal across all baselines in every time and frequency bin. A lightcurve can then be derived by taking a mean of $S(t,\nu)$ over the frequency axis. Appropriate smoothing can be applied to suppress noise, and emphasize variability on particular time and frequency scales.

This approach is computationally expensive, scaling as $O(N_\mathrm{vis}\times N_\mathrm{src})$, and is thus only practical for relatively few sources. We will apply it to the PARROT below. An alternative is to leverage the FFT to compute dynamic spectra \emph{en masse}: essentially, a dirty image cube made from the ultimate residuals, with appropriate time and frequency binning, will contain a dynamic spectrum at every image pixel (an alternative way to think about it is that the beamforming procedure above is mathematically equivalent to what the Fourier transform does at every pixel in the image).
This scales as $O(N_\mathrm{vis}\times N_\mathrm{pix} \log N_\mathrm{pix})$. 

The dirty image cube approach is very easy to implement (only a run of the imager is needed, no deconvolution) and is embarrassingly parallel over the time and frequency bins. In practice, it is only limited by considerations of RAM and disk space -- 4D images get prohibitively large very quickly. A compromise is to reduce the number of time/frequency bins. At the extreme, using one frequency bin corresponding to the full band, the dirty cube becomes a 3D cube of lightcurves. This is the approach we take in this work.

\subsection{Lightcurve extraction pipeline}
\label{sec:lightcurves}

The lightcurve extraction pipeline is a continuation of the imaging pipeline described in Sect.~\ref{sec:followup-pipeline} (and, in fact, forms part of the same \stimela\ recipe). We therefore retain the step numbering from Sect.~\ref{sec:followup-pipeline} in what follows.

\vspace{1ex}
\noindent{\bf Step 6. A source catalog} is constructed by running the \pybdsf\footnote{\url{https://github.com/lofar-astron/PyBDSF}} tool \citep{pybdsf} on the images produced by step 3 of the pipeline. This results \revone{on the order of} 23,000 source detections in the L-band images, and 42,000 detections in the UHF-band images. Note that this is a one-off operation that is not required for every pipeline run (or for every observation); it is therefore skipped by default and only invoked by hand. It is, in any case, preferable to have a single catalog to work with across multiple observations, so as to retain a consistent set of source labels. We also use \pybdsf\ to get in-band spectral index estimates, needed for the source selection below. This requires power beam-corrected per-subband images. We use the MeerKAT holography beams provided by \citet{mdv-beams} to correct the step 3 images by the array-average power beam, and make a spectral cube suitable for use by \pybdsf.

Although the following steps can (and do) routinely produce lightcurves for any number of sources, analysing them at such scale is well beyond the scope of this work. A sensible selection is needed to narrow down the sources of interest (in addition to the PARROT). Since only extremely compact sources can be expected to show variability on such timescales (including due to scintillation), we perform the following selection:

\begin{description}
    \item[{\bf AGN core candidates:}] unresolved sources with a flat or inverted spectrum. The L-band catalog produced at step 6 contains 26 sources with an apparent flux of 0.3 mJy or higher, and a spectral index of -0.1 and above. 
    \item[{\bf Compact VLBI sources:}] the Radio Fundamental Catalog\footnote{\url{http://astrogeo.org/rfc/}} contains 9 sources within our (UHF) field of view. We find bright counterparts to all 9 sources in the UHF catalog from step 3 (with positions matching to within 
    $0.5\sim4\farcs$).
    \item[{\bf A known pulsar,}] PSR J2012-2029 is approximately 45$\arcmin$ East of the field centre, and has a reported continuum flux density of 1 mJy at 800 MHz \citep{psr2012}. Its uncertainty in Ecliptic latitude is probably quite large, because the position is derived from pulse timing and the source has a very low Ecliptic latitude. In the discovery paper \citep{psr2012} the error ellipse was described by its projection onto Equatorial coordinates; in Ecliptic coordinates the uncertainty may be as large as $8\arcmin$ in latitude, and $4\farcs5$ in longitude. The UHF catalog from step 6 contains three sufficiently bright candidates within a 3$\sigma$ error box.
\end{description}

The selected candidate positions are stored as DS9 region files, for subsequent use by the lightcurve extraction tool.

\vspace{1ex}
\noindent{\bf Step 7. Lightcurve cube.} High time cadence (HTC) dirty images are constructed from the ultimate residual visibilities, by running \wsclean\ with the {\tt -intervals-out} parameter set to the number of available integrations. This results in several thousand images (one per each 8s integration) per each observation, which we then stack into a single 3D FITS cube. To keep this cube to a reasonable size, and because oversampling the PSF is not required (as we will not be doing deconvolution), we set the pixel size to 3 times that used in the regular imaging pipeline, that is $2\farcs4$ for L-band and $4\farcs8$ for UHF, and set the image size to $3072\times3072$.

\vspace{1ex}
\noindent{\bf Step 8. Cube smoothing:} the lightcurve cube is convolved with a Gaussian filter along the time axis, to increase SNR and emphasize timescales of interest. This step is optional -- even the raw 8s lightcurves already show interesting events. It can be repeated several times with different kernel widths ($\sigma_t$ values) to obtain lightcurves of various degrees of smoothness. 

\vspace{1ex}
\noindent{\bf Step 9. Light curve extraction:} a Python script loads the (raw or smoothed) cube and the source catalog, as well as the DS9 regions corresponding to ``interesting'' sources. For every source in the catalog (beginning with the RATT, followed by the ``interesting'' sources, then the remaining sources in order of decreasing flux), it identifies the pixel position nearest to the source position, and extracts a slice through the time axis at that pixel. It also estimates the image rms in a box surrounding the pixel, at each timeslice, and reports this as the associated error bar. The results are stored to disk, and plotted for inspection.

\newlength{\lcwidth}
\setlength{\lcwidth}{.91\textwidth}

\newcommand{\lc}[1]{%
\begin{minipage}{\lcwidth}%
\includegraphics[width=\lcwidth]{#1}%
\end{minipage}\vspace{1em}%
}

\begin{figure*}
\begin{tabular}{@{}cc@{}}
L0&\lc{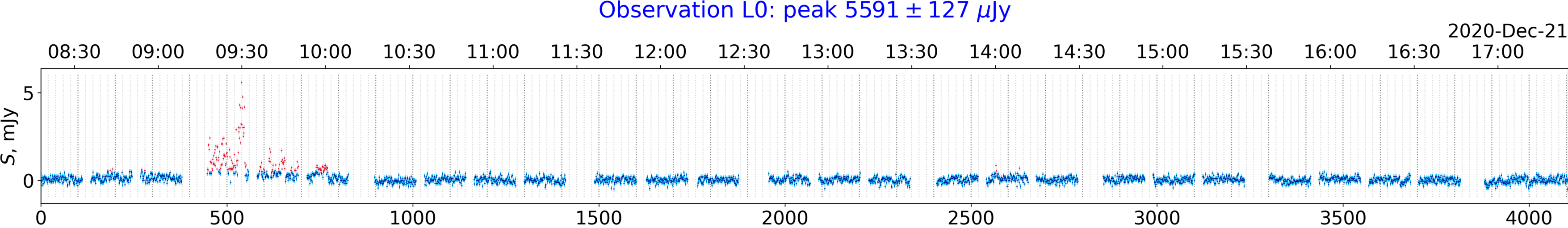}\\
L1&\lc{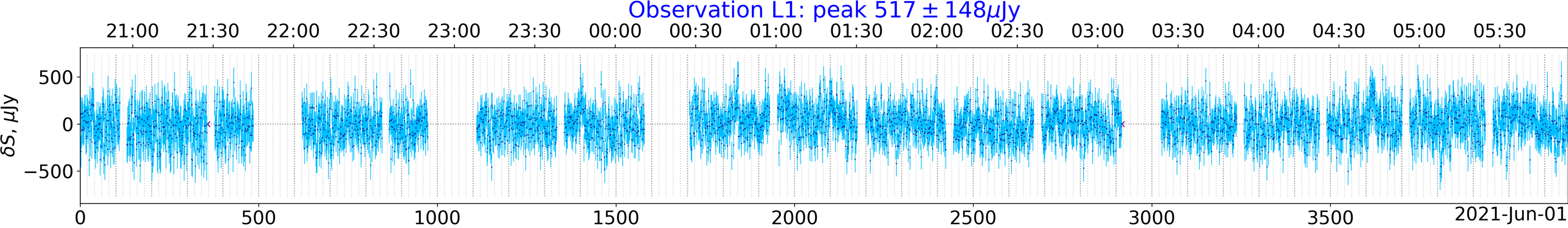}\\
L2&\lc{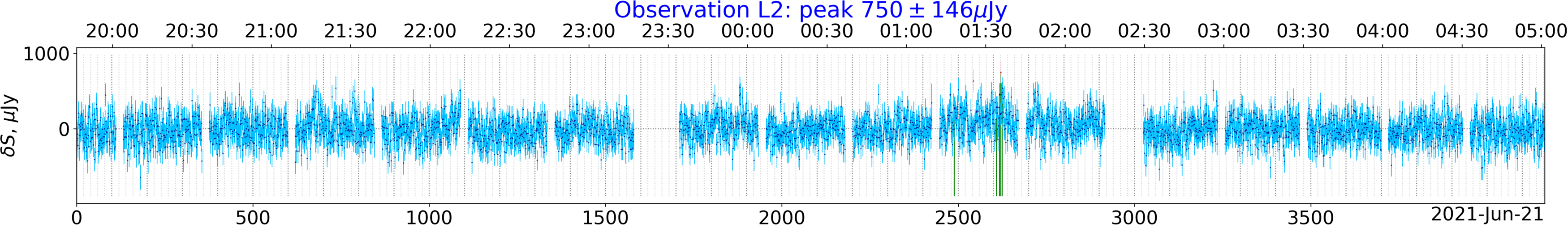}\\
L3&\lc{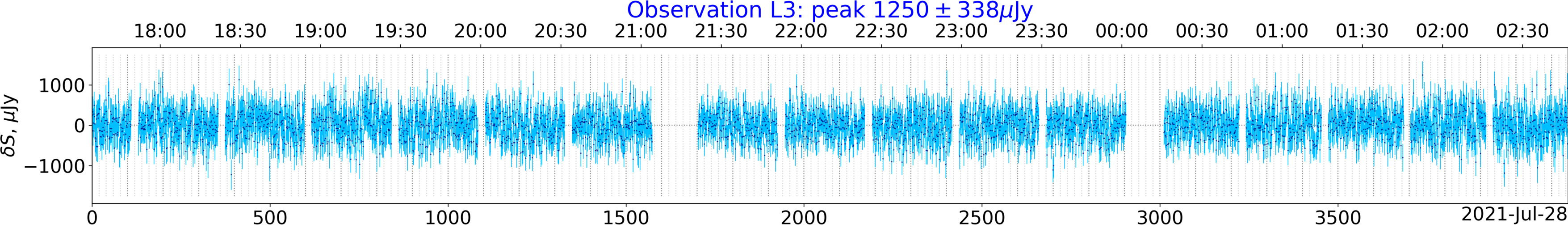}\\
U0&\lc{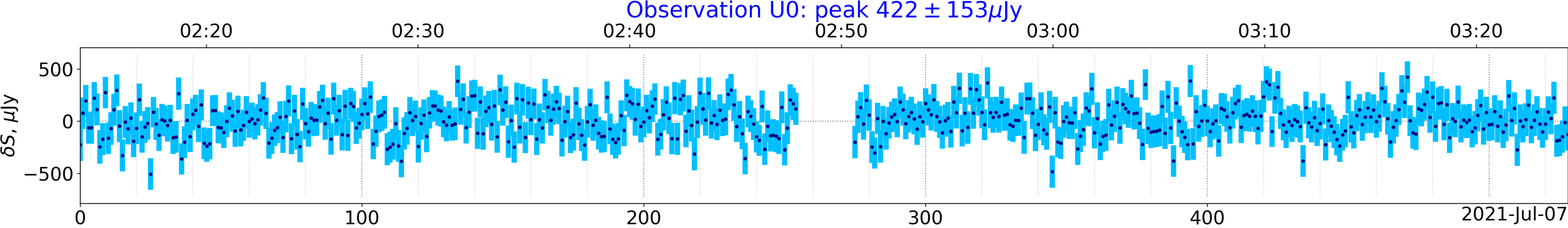}\\
U1&\lc{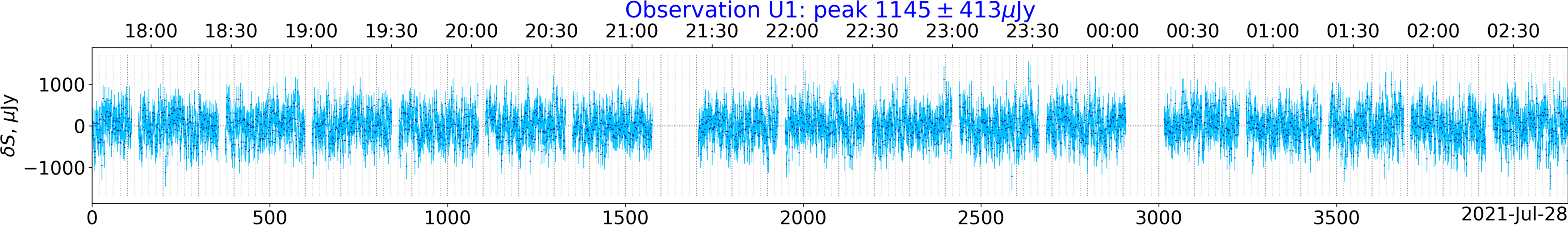}\\
U2&\lc{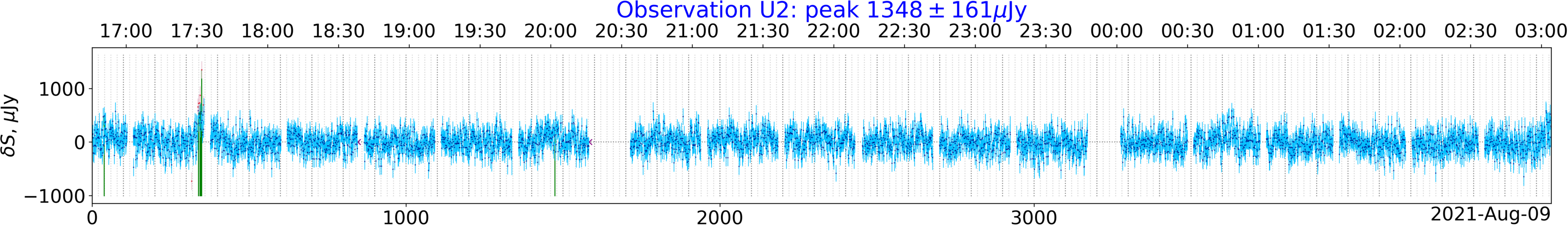}\\
U3&\lc{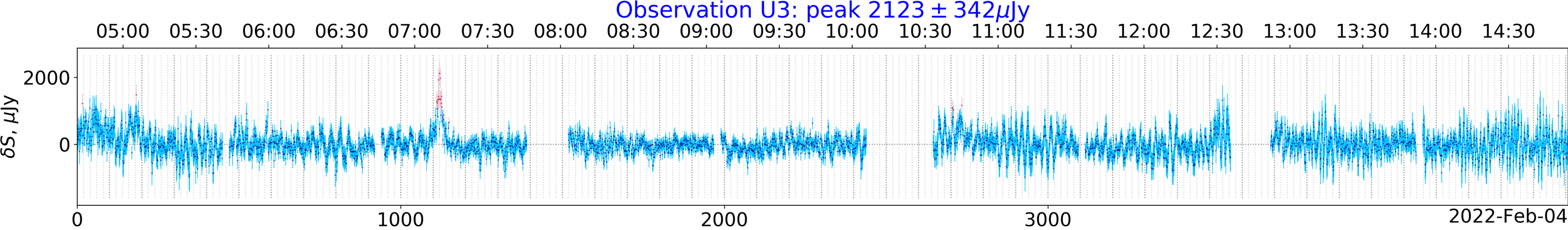} 
\end{tabular}
\caption{\label{fig:lc-rrat-1}
Lightcurves of the PARROT (mean-subtracted, for all observations except L0) at raw 8s sampling, from all available observations (Table~\ref{tab:obs}). 
Recall that observations L3 and U1 were carried out simultaneously in split-array mode, and that U0 was short.
Four-sigma excursions are plotted in red (with light red error bars), all other measurements in blue (light blue error bars). 
The vertical green bars indicate MeerTRAP pulse detections. Bottom X axis is timeslot number, top X axis is time (UTC).
}
\end{figure*}

\begin{figure*}
\begin{tabular}{@{}cc@{}}
L0&\lc{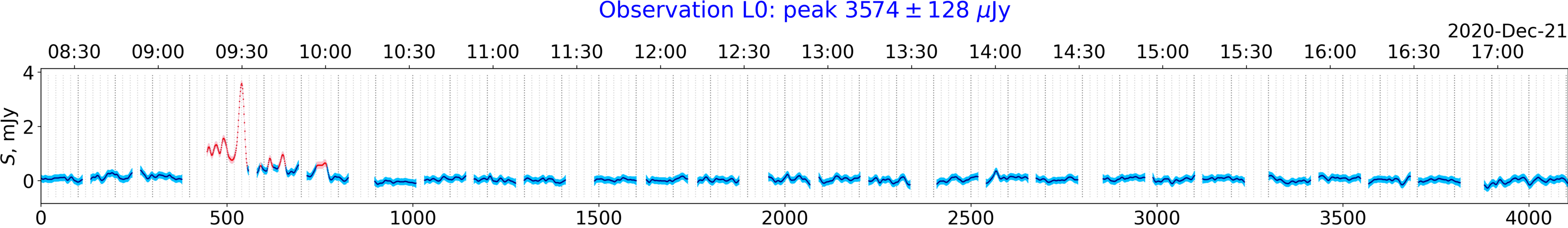}\\
L1&\lc{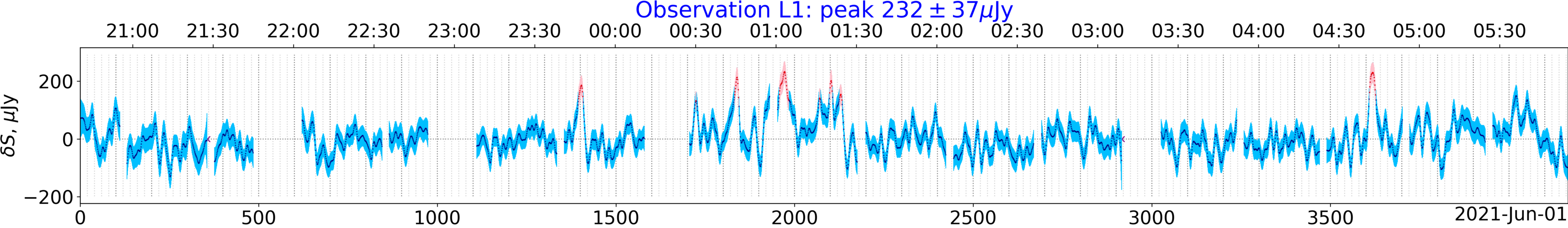}\\
L2&\lc{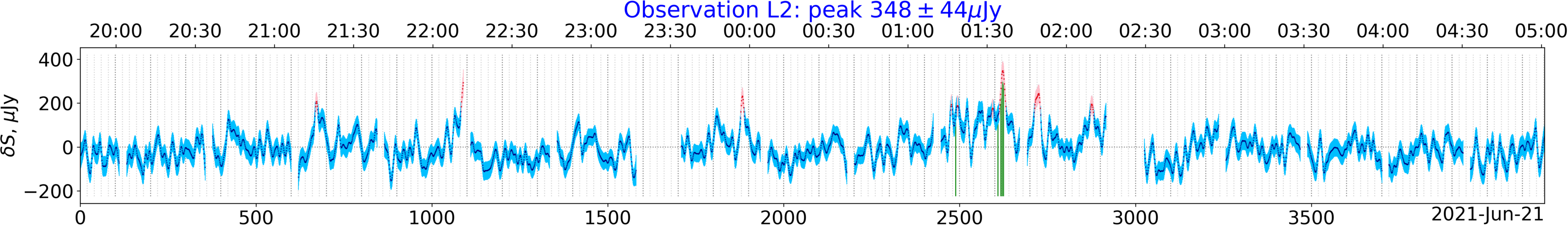}\\
L3&\lc{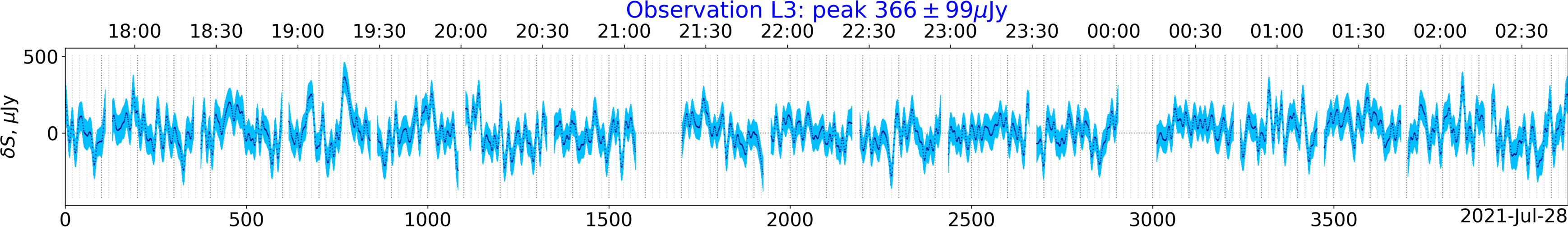}\\
U0&\lc{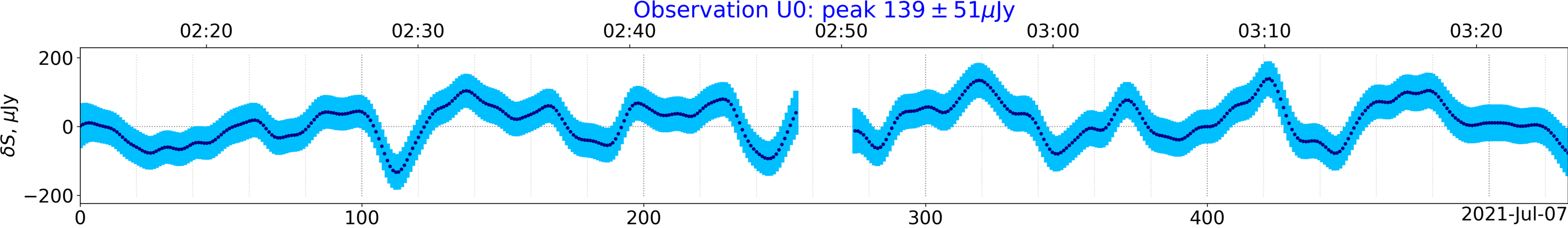}\\
U1&\lc{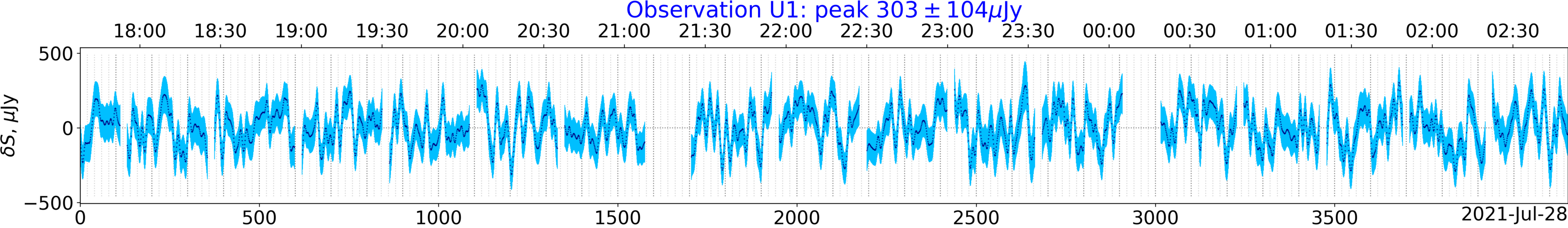}\\
U2&\lc{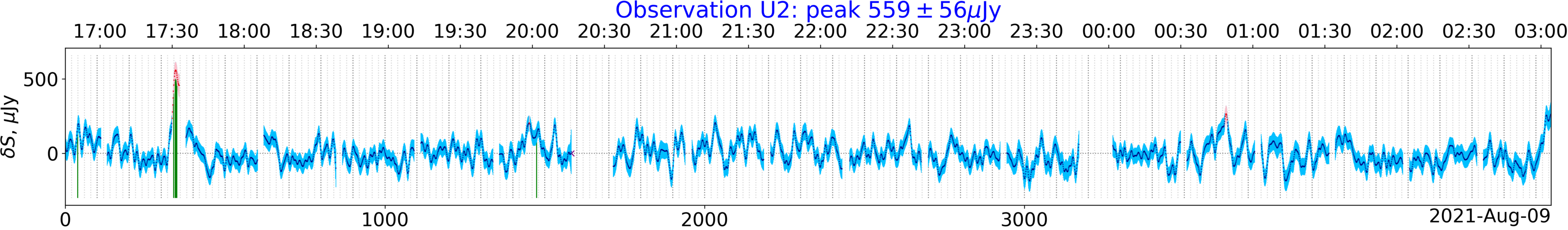}\\
U3&\lc{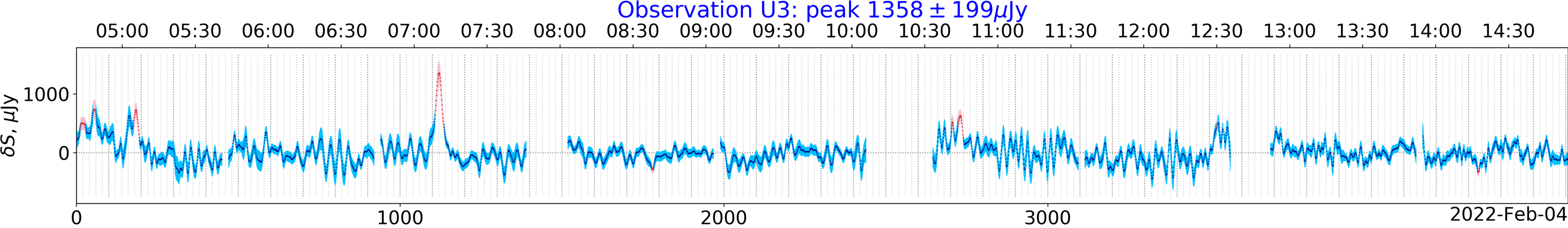} 
\end{tabular}
\caption{\label{fig:lc-rrat-16}
Lightcurves of the PARROT (mean-subtracted, for all observations except L0) smoothed with a 30s Gaussian filter, from all available observations (Table~\ref{tab:obs}). Layout and colour coding is as per previous figure.}
\end{figure*}

\begin{figure*}
\includegraphics[width=\lcwidth]{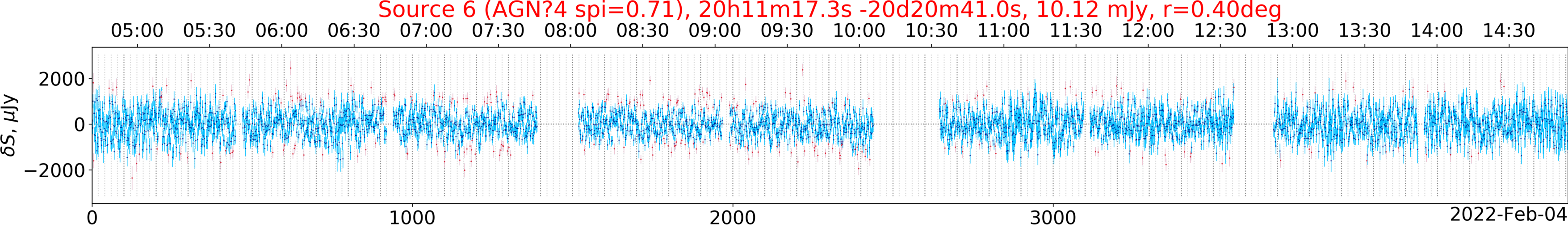} \\
\includegraphics[width=\lcwidth]{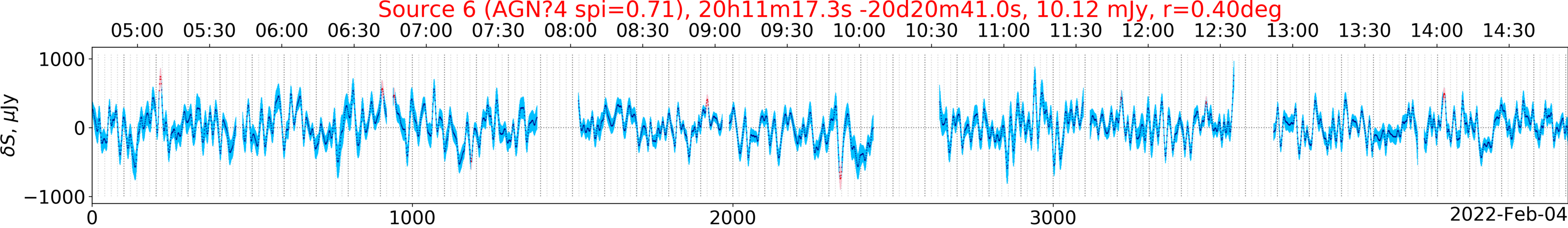} \\
\includegraphics[width=\lcwidth]{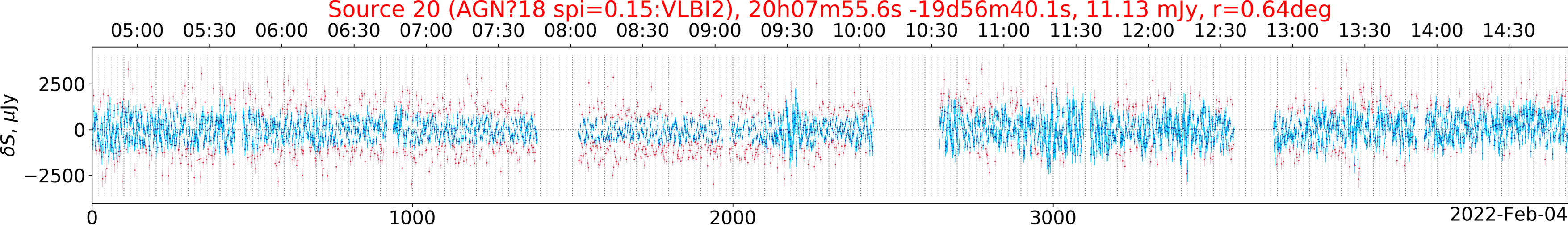} \\
\includegraphics[width=\lcwidth]{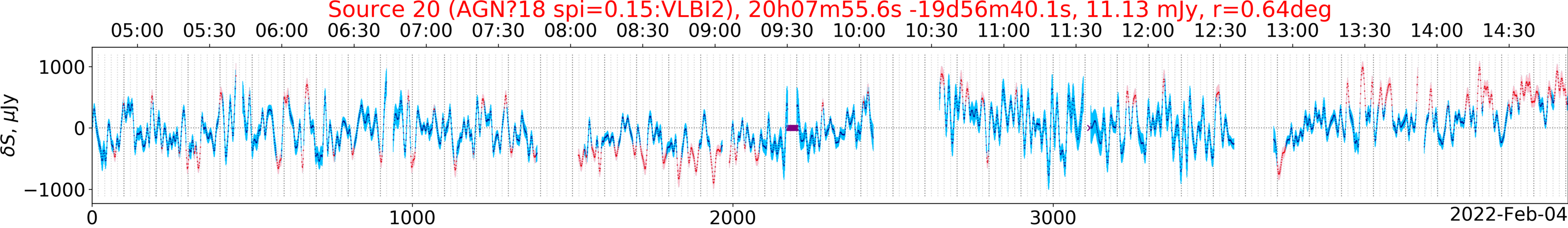} \\
\includegraphics[width=\lcwidth]{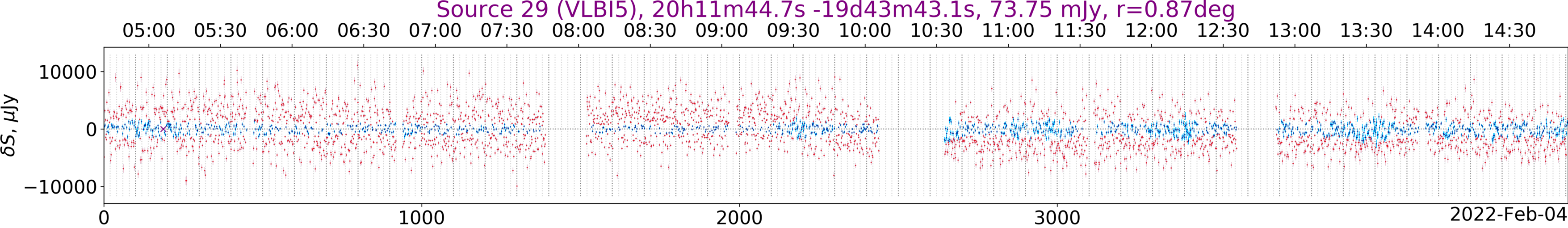} \\
\includegraphics[width=\lcwidth]{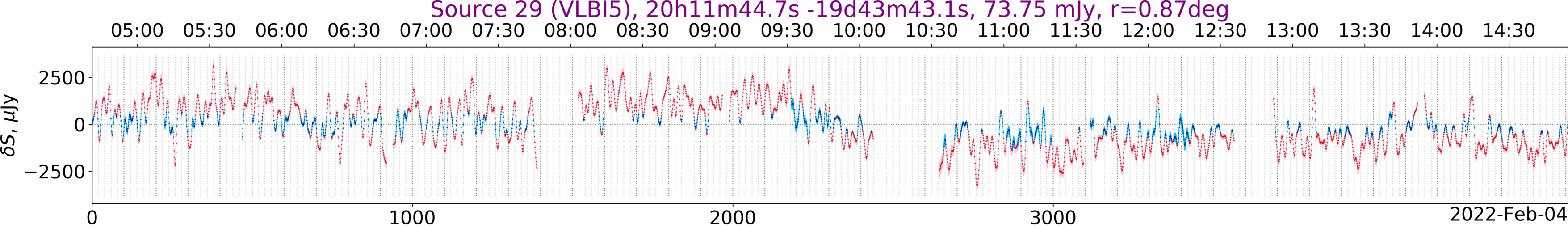} 
\caption{\label{fig:lc-u}
Mean-subtracted lightcurves for three selected other sources from observation U3. Alternating rows show lightcurves at the raw 8s sampling, and smoothed with a 2-minute Gaussian filter. Colour coding and axis layout is as per previous figures.}
\end{figure*}

\begin{figure*}
\includegraphics[width=\lcwidth]{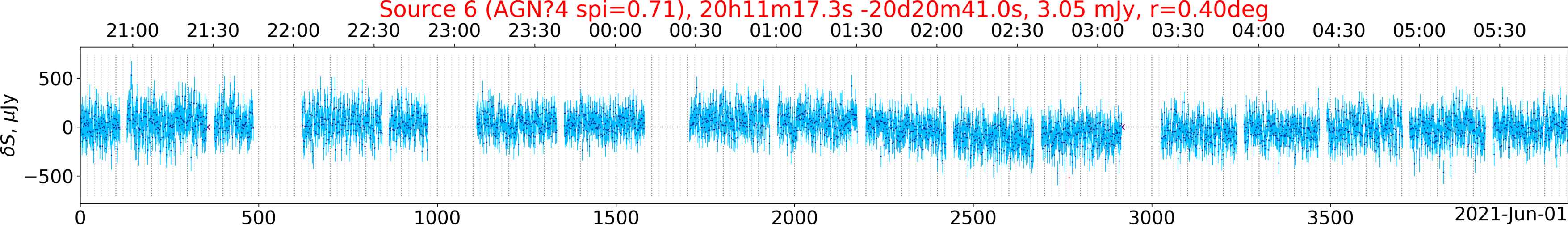} \\
\includegraphics[width=\lcwidth]{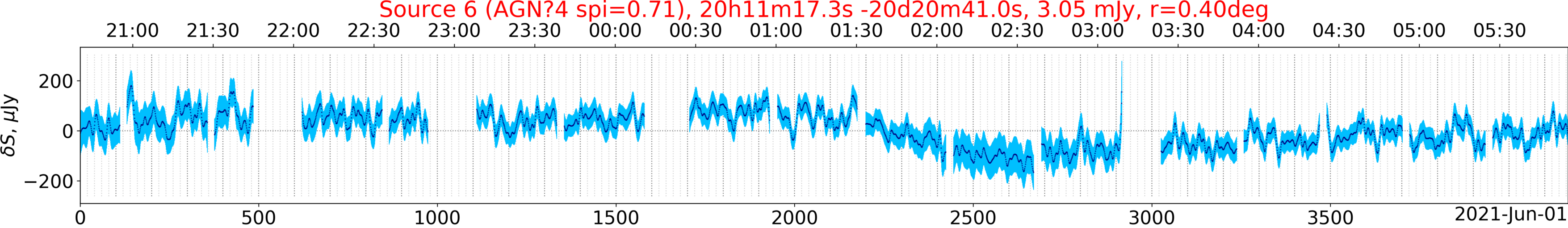} \\
\includegraphics[width=\lcwidth]{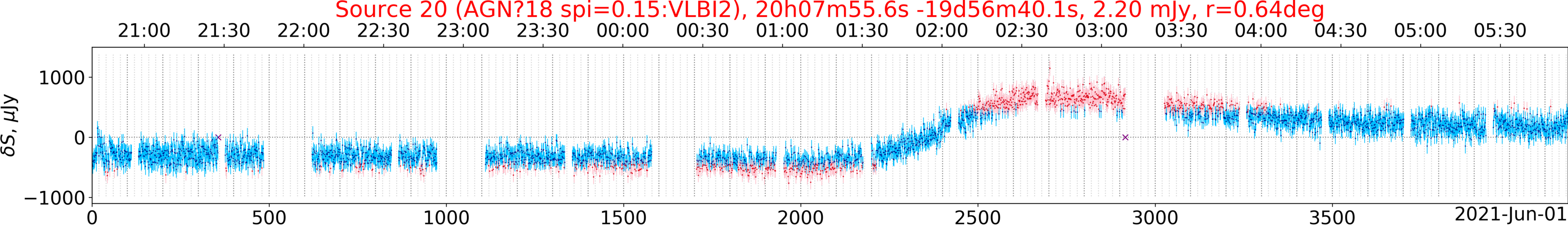} \\
\includegraphics[width=\lcwidth]{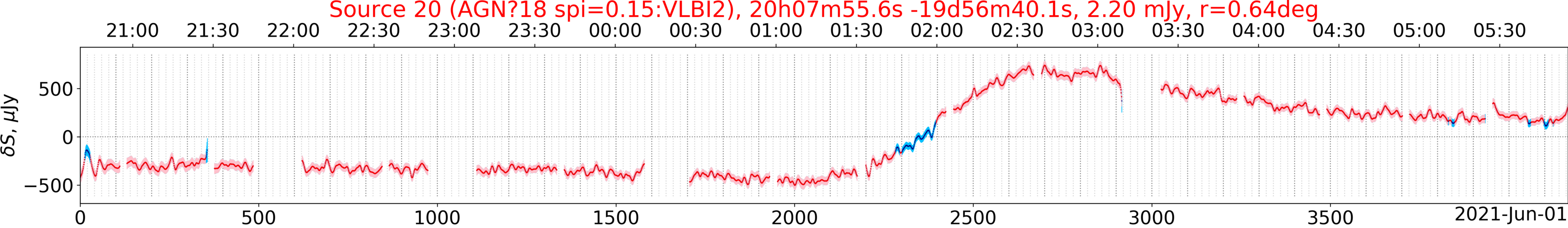} \\
\includegraphics[width=\lcwidth]{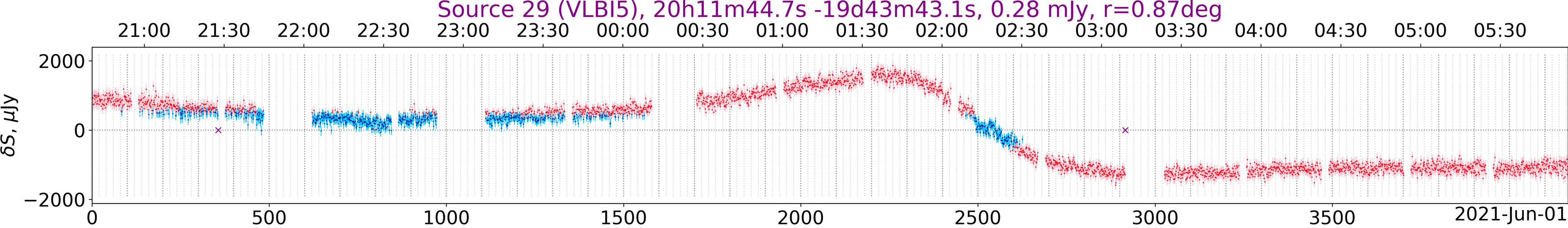} \\
\includegraphics[width=\lcwidth]{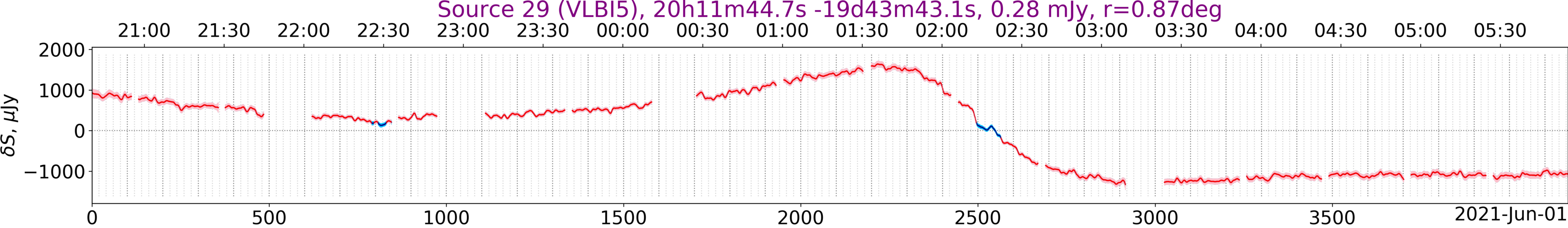} 
\caption{\label{fig:lc-l}
Mean-subtracted lightcurves from observation L1, for the same sources as in Fig.~\ref{fig:lc-u}. Alternating rows show lightcurves at the raw 8s sampling, and smoothed with a 2-minute Gaussian filter. Colour coding and axis layout is as per previous figures.
}
\end{figure*}

\subsection{Dynamic spectra extraction}

\label{sec:dynspectra}

\def\BL{{(pq),t\nu}}
\def\BLpq{{\vec{b}^{t\nu}_{pq}}}
\def\BL{{\vec{b}}}
\def\Vres{{\bm \delta}^{\textbf{v}^{c}}_\BL}
\def\LOTSS/{{\sc LoTSS}}
\def\LoTSS/{{\sc LoTSS}}
\def\RIME/{RIME}
\def\PSF/{{\sc psf}}
\def\bl{\footnotesize\vec{b}}
\newcommand{\conj}[1]{\overline{#1}}
\renewcommand{\vec}[1]{\mathbf{#1}}
\def\DDE/{{\sc dde}}

In this section we briefly describe the mathematical underpinnings of
the {\tt DynSpecMS} tool\footnote{\url{https://github.com/cyriltasse/DynSpecMS}}, which
we used to extract the dynamic spectra. The full approach (taking into account other effects such as the \DDE/s)
will be presented in Tasse et al. (in prep), in the context of the exploitation of the LOFAR 
Two-Meter Sky Survey (LoTSS) data.

The inverse \RIME/ states that each synthesised beam (in the dirty or
residual image) can be viewed as the weighted sum of all
visibilities. However, imaging all pixels for all times and
frequencies in typical observation would require estimating a large number
of pixels $N_\nu \times N_t \times N_{\mathrm{pix}}$, making this impractical.
To address this issue, we select the few to hundreds directions of
interest $N^{\mathrm{sel}}_{\mathrm{pix}}$ and image $N\nu\times N_t \times
N^{\mathrm{sel}}_{\mathrm{pix}}$ pixels, using a Direct Fourier Transform (DFT). This
approach is also used and described by \citet{Zic19}. Conceptually, this is similar
to Multi-Object Spectroscopy used in the optical and near-infrared
energy domains, where only a limited number of fibers are put in the
focal plane of telescopes to acquire spectra. This technique can be
extended to dynamic spectro-polarimetry, estimating the Stokes $IQUV$
parameters for all available times and frequencies. 

One challenge of image synthesis with interferometric data, as opposed
to optical telescopes, is the sparse coverage of the uv-domain. This
means that images built by Fourier-transforming the visibilities in
the uv-domain must be deconvolved to interpret flux densities. The
\PSF/ is non-zero even far from its origin, and sources can
cross-contaminate each other. This is especially true when synthesizing images over
narrow time-frequency domains with poor uv-coverage. This leads to
pixels in a given direction in the time-frequency space having
contributions from all time-frequency sidelobes of all sources in the
field. To address this, we work exclusively with residual
visibilities and assume that the majority of
sources that exhibit variation in time-frequency space are compact,
allowing us to interpret the peak flux in the residual image as a
physical integrated flux density. Specifically, the residual
corrected visibilities $\Vres$ on baseline
$\BL\leftrightarrow\{pqt\nu\}$ can be written as

\begin{alignat}{2}
\label{eq:ResVis}
\Vres =&\textbf{v}^{c}_\BL-
\int_{\vec{s}}
\textbf{x}_{\vec{s}\nu}
k^{\vec{s}}_{\BL} \textrm{d}\vec{s}\\
\text{with }k^{\vec{s}}_\BL=&\exp{\left(-2 i\pi \frac{\nu}{c}\vec{b}_{pq,t}^T(\vec{s}-\vec{s}_0)\right)}
\end{alignat}

\noindent where $\textbf{v}^{c}_\BL$ is the corrected visibility on
baseline $\BL$, $k^{\vec{s}}_{\BL}$
 is the
geometric phase term of the \RIME/ and
$\widehat{\textbf{x}_{\vec{s}\nu}}$ is the sky model. The polarimetric dynamic spectrum
$\textbf{x}_{\vec{s},t\nu}$ in a given direction $\vec{s}$ is then
built as

\def\Dxtnu{{\bm \delta}^{\textbf{x}}_{\vec{s},t\nu}}

\begin{alignat}{2}
\label{eq:DynSpec}
\widehat{\textbf{x}_{\vec{s},t\nu}}=&\widehat{\textbf{x}_{\vec{s},\nu}}+{\bm \delta}^{\textbf{x}}_{\vec{s},t\nu}\\
\text{with } \Dxtnu=&
\displaystyle\sum\limits_{(pq)}
        {\bm \delta}^{\textbf{v}^{c}}_\BLpq
\conj{k^{\vec{s}}_\BLpq}
\end{alignat}

\noindent where $\conj{k}$ is the complex conjugate of $k$, and the term $\Dxtnu$ is an 
$IQUV$-pixel for a given
direction $\vec{s}$, time $t$ and frequency $\nu$. It is the sum of all
visibilities phased in that direction.

The noise in the dynamic spectra has a complex structure and can
significantly vary from one $(t,\nu)$-coordinate to another. This is
mostly due to (i) the flagging fraction being very different across
time and frequency. Assuming a known variance per visibility, we can
have a variance dynamic spectra. But (ii) since we work with corrected
data, the variance of the individual visibility is hard to estimate,
and systematic effects are seen even after correcting for (i). So
instead of estimating the variance assuming a statistical model, we
directly measure it from the data. To do this, a fixed number of
dynamic spectra are synthesized at ``off'' positions selected at random
within a $1\degr$ radius, and the variance of the dynamic spectra is measured
from these realizations using the Median Absolute Deviation estimator.

\begin{figure}
    \includegraphics[width=0.48\textwidth]{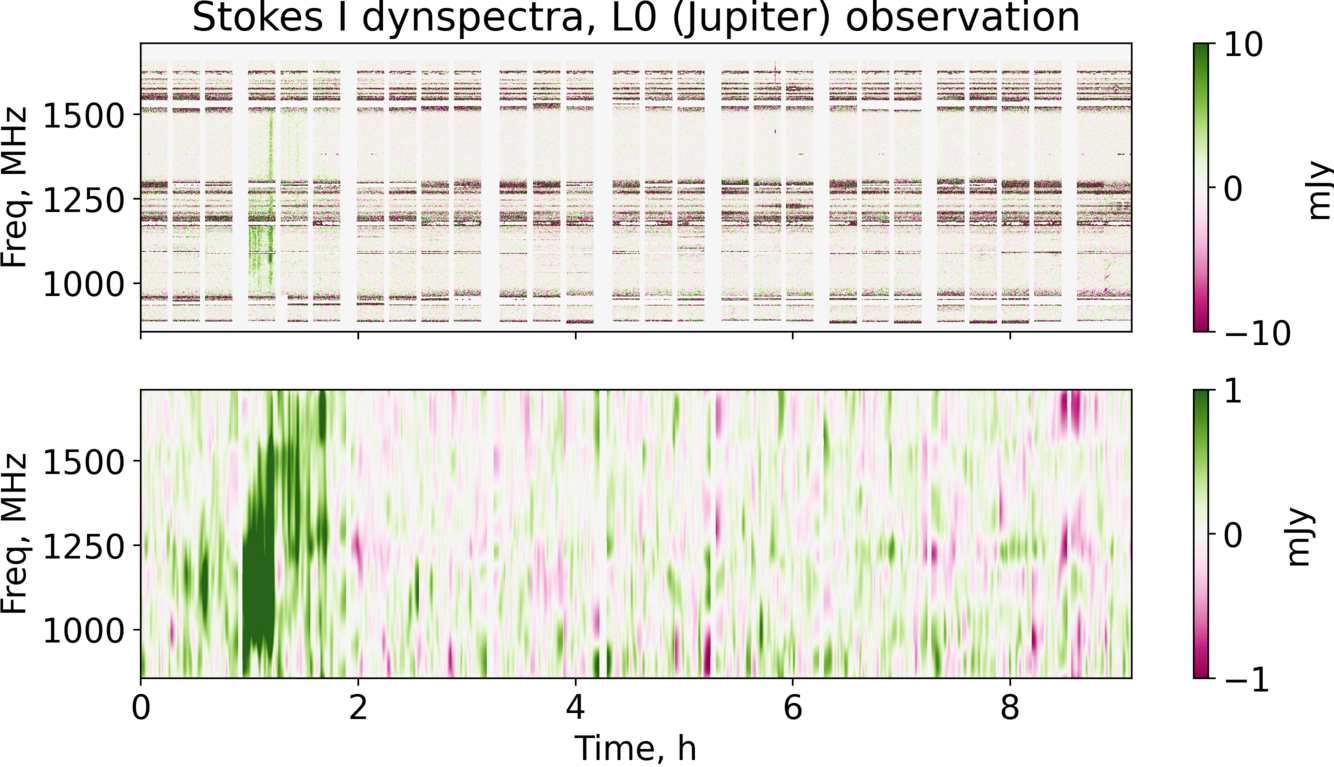}
    \caption{\label{fig:ds-jup-i}
    Raw and Gaussian-smoothed dynamic spectra of the PARROT in Stokes $I$, L0 (Jupiter) observation.}
\end{figure}
\begin{figure}
    \includegraphics[width=0.48\textwidth]{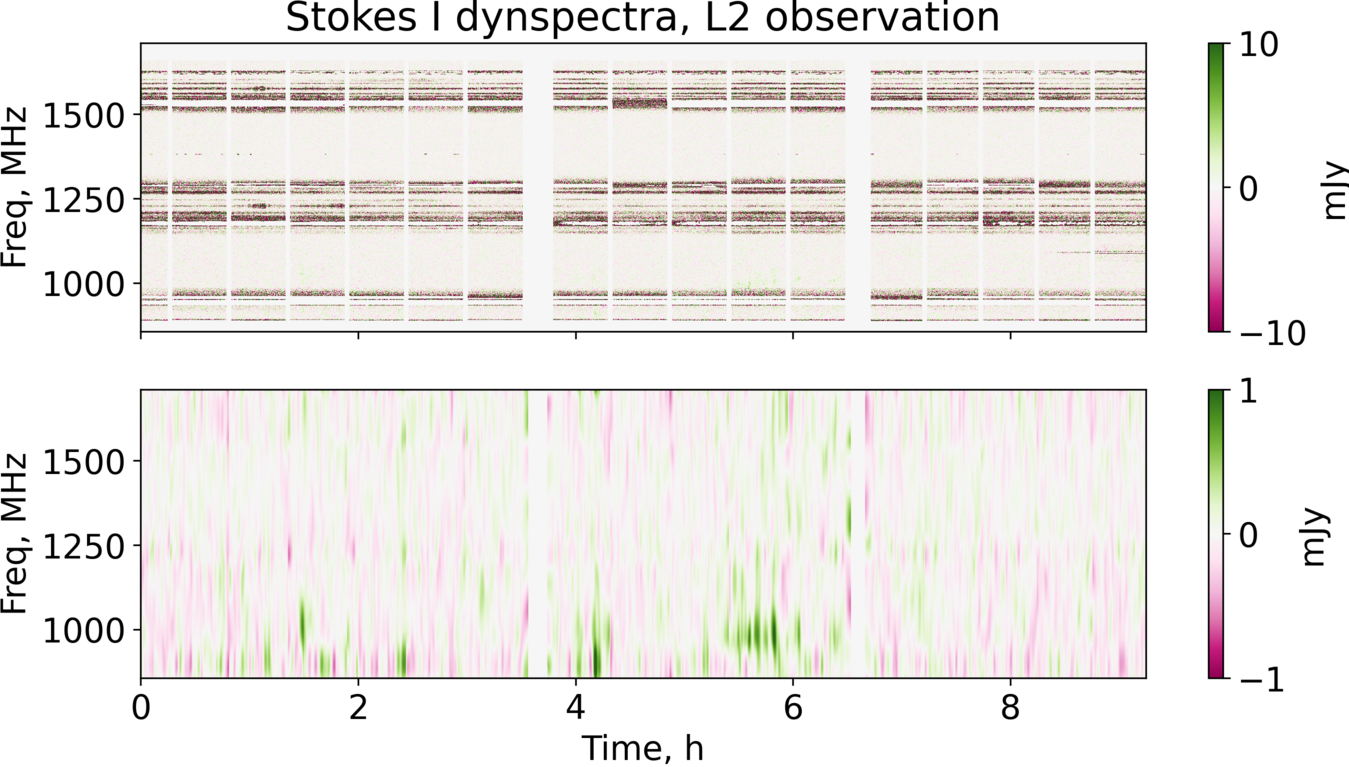}
    \caption{\label{fig:ds-l2}
    Raw and Gaussian-smoothed mean-subtracted dynamic spectra of the PARROT in Stokes $I$, L2 observation. \revone{Note the several clear detections just before and after 2h, and between 4 and 6h.}}
\end{figure}
\begin{figure}
    \includegraphics[width=0.48\textwidth]{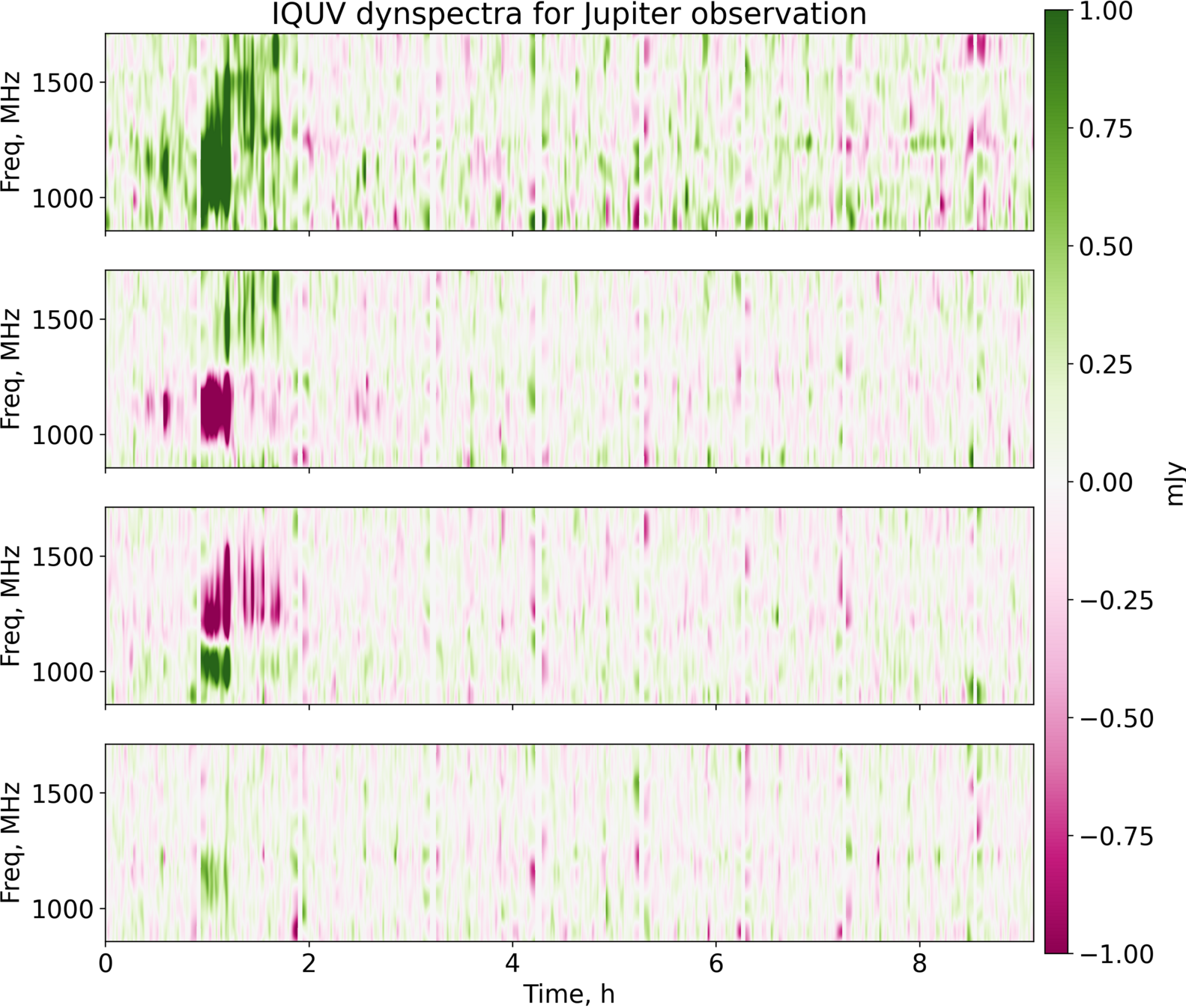}
    \caption{\label{fig:ds-jup-pol}Gaussian-smoothed dynamic spectra of the PARROT in Stokes $IQUV$, L0 (Jupiter) observation.}
\end{figure}

\setlength{\plotheight}{0.14\textwidth}

\begin{figure}
\begin{tabular}{@{}l@{}}
\includegraphics[height=\plotheight]{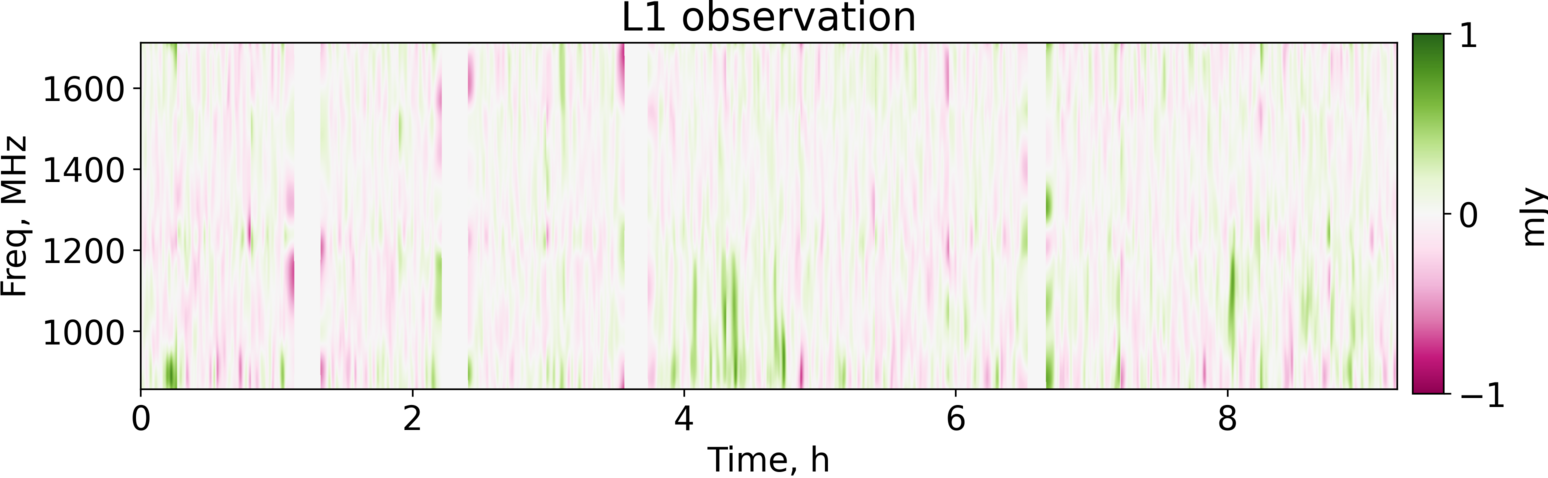}\\
\includegraphics[height=\plotheight]{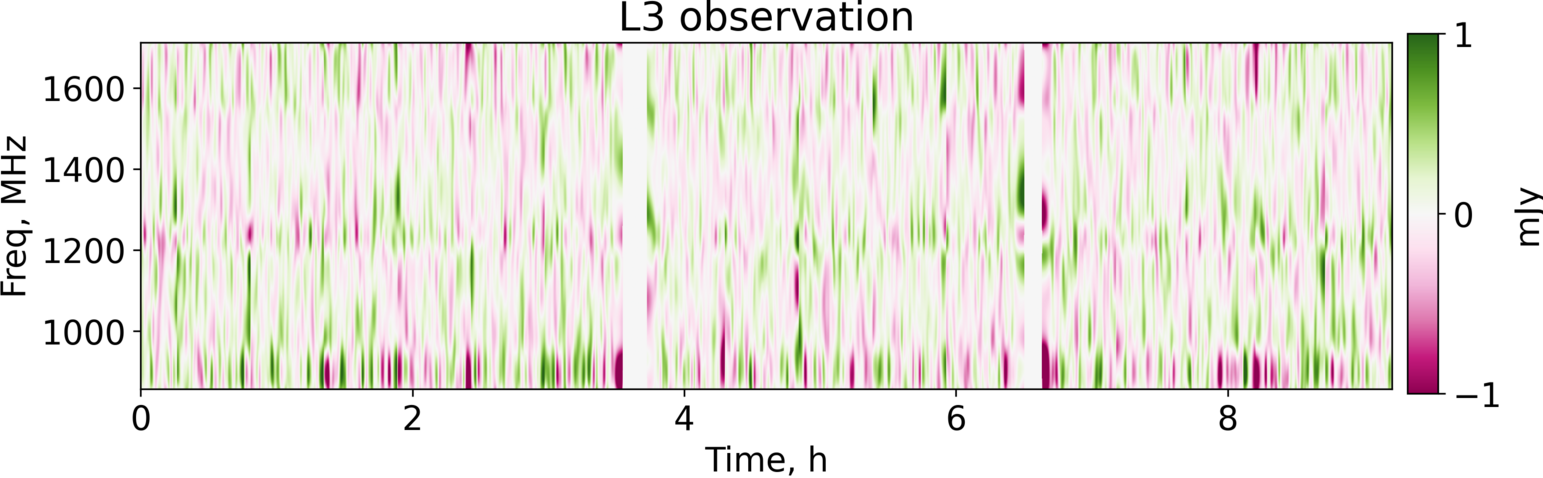}\\
\includegraphics[height=\plotheight]{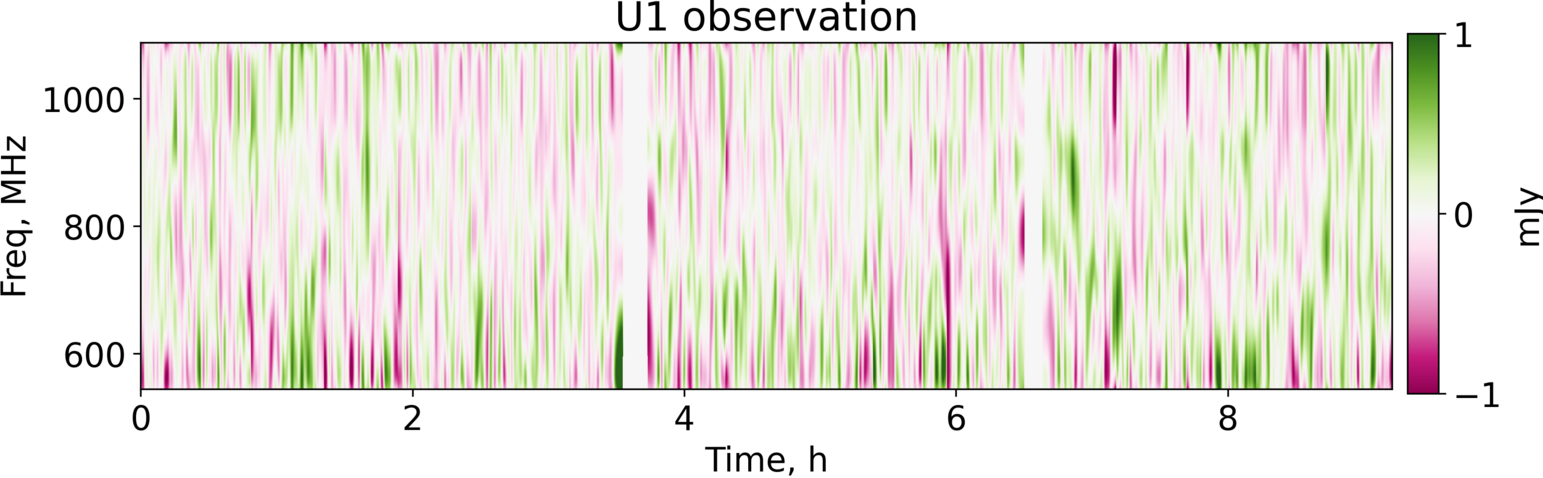}\\
\includegraphics[height=\plotheight]{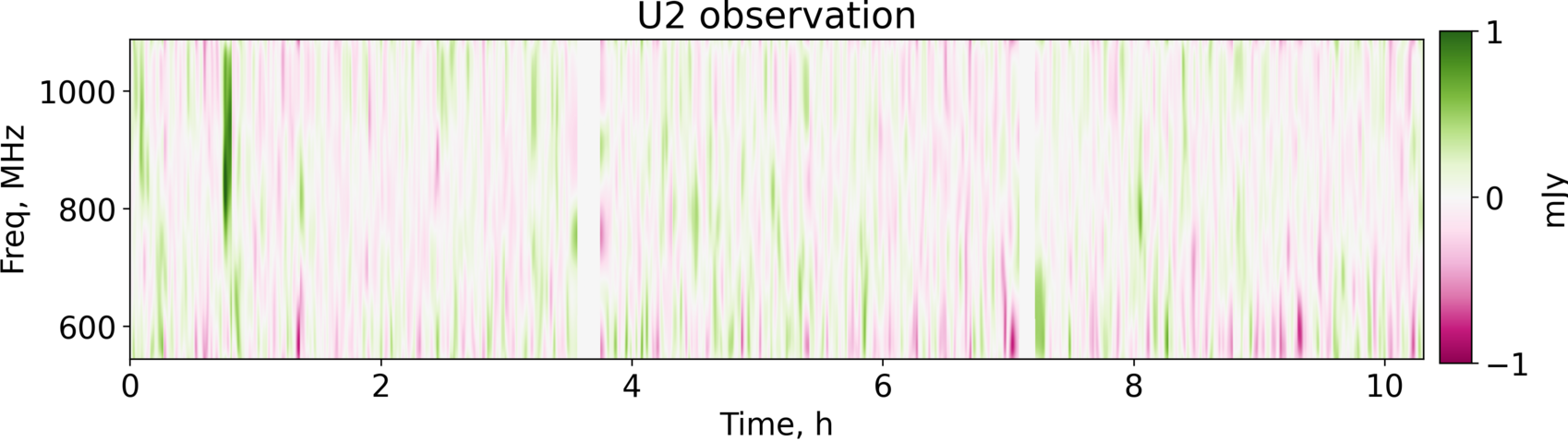}\\
\includegraphics[height=\plotheight]{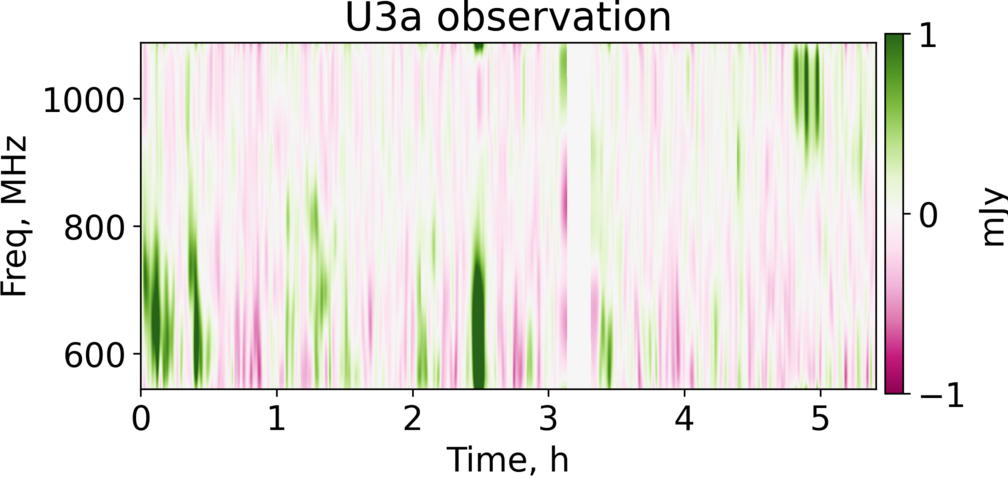}\\
\includegraphics[height=\plotheight]{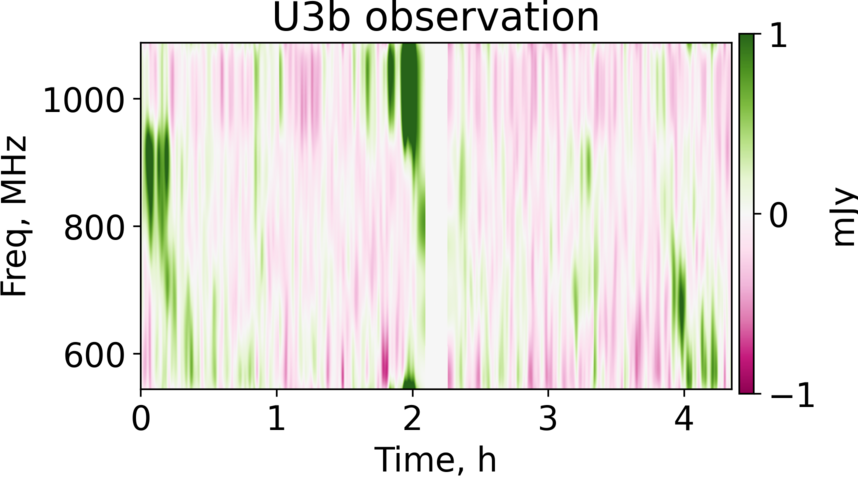}
\includegraphics[height=\plotheight]{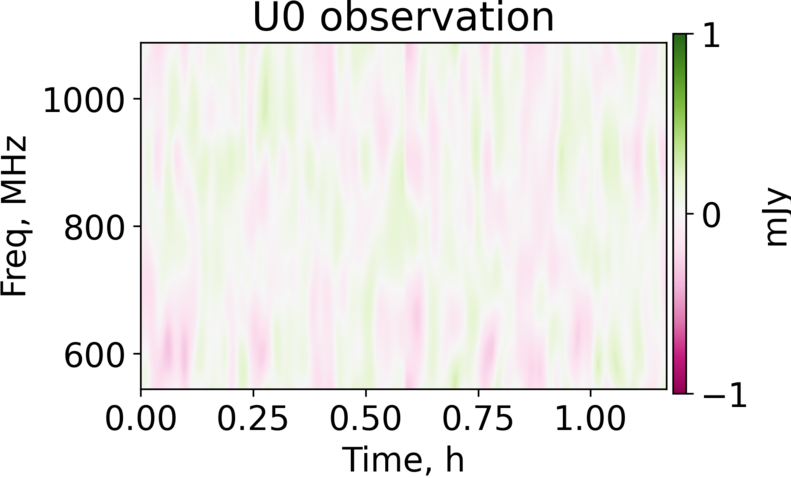}
\end{tabular}
\caption{\label{fig:ds-others}
Gaussian-smoothed, mean-subtracted dynamic spectra of the PARROT in Stokes $I$ for other follow-up observations. Note 
detections in L1 and U2 (the latter coincident with a MeerTRAP pulse detection), and many strong detections in U3a/b. 
}
\end{figure}

\section{MeerTRAP detections and Pulsed Dynamic Spectra}
\label{sec:timing}

The detection of 5 single pulses during the L2 observation of the transient source established its nature as a radio emitting neutron star, PSR\,J2009$-$2026 a.k.a. PARROT. The single pulses were detected using the Meer(more) TRansients and Pulsars (MeerTRAP) pipeline \citep{msb+20,rsw+21} which is a real-time commensal mode backend running on MeerKAT observations. Details of how MeerTRAP is used to find Galactic sources are given in \cite{bbc+22} and we will only summarise them here. Data from the MeerKAT F-engine (which channelises the individual receiver data) are collected by the Filterbank BeamFormer User Supplied Equipment (FBFUSE) cluster \citep[][]{2021chen} and it forms up to 768 coherent beams (CBs) on the sky, typically using about 40 dishes in the dense core region of MeerKAT. An incoherent beam (IB) is also formed using the sum of the data from up to 64 MeerKAT antennas. The beamformed data are transferred over the network to MeerTRAP's computer cluster, Transient User Supplied Equipment (TUSE). For these observations FBFUSE combines the data in frequency and time to give 1024 frequency channels, independent of observing band, and a sampling time of approximately 482~$\mu$s in the UHF band and approximately 306~$\mu$s in the L-band. These data are then searched in real-time over a range of dispersion measures (DM) from 20~pc~cm$^{-3}$ (this lower limits is chosen to minimise the contributions from terrestrial RFI near dispersion measure of zero) up to a maximum of either 2500~pc~cm$^{-3}$ (UHF-band) or 5000~pc~cm$^{-3}$ (L-band) for pulses with S/N greater than seven (for these observations, this was later increased to eight). 

MeerTRAP was not included in the original conjunction observation and did not detect any pulses during the L1 run when the transient was first placed at the phase centre. The change to the phase centre means that for this, and all future observations with MeerTRAP, the source would be expected to be seen in the first coherent beam (CB0). The pulses detected in L2 were found to have DMs in the range 21--27~pc~cm$^{-3}$, the reason for the wide range is likely due to the influence of the zero-DM filter \citep{2009eatough} that is applied to the data and can sometimes distort low-DM bright pulses. This may have also prevented us from detecting other pulses if the DM dropped below our DM threshold of 20~pc~cm$^{-3}$, or reduced the S/N too much. The pulses were found in an interval of approximately 1100\,s with the first at UT 01:18 and the the remaining four in a two minute period centred on UT 01:35 (see Figs.~\ref{fig:lc-rrat-1} and \ref{fig:lc-rrat-16}). This distribution of pulse arrival times allowed us to establish a period of 1.638(5)\,s. 

No pulses were detected during U0 and MeerTRAP did not run during the L3/U1 observation. However in the U2 observation a total of 12 pulses were detected in the UHF band. The times of arrival of the pulses are again shown in 
Figs.~\ref{fig:lc-rrat-1} and \ref{fig:lc-rrat-16}, with 10 of the 12 pulses happening in a 77\,s interval that corresponds to the brightest epoch of U2 most clearly seen in Figure \ref{fig:lc-rrat-16}.  The lower frequency of the UHF band detections allowed us to determine an improved DM, but it is still constrained by the typical pulse width of around 50\,ms and the sometimes double peaked nature of the pulses. This suggests that the DM is closer to 20~pc~cm$^{-3}$, confirming the suspicion that some pulses may have been missed as they fell below our DM cutoff. 

Having confirmed that the source is radio emitting neutron star of some type for observation U3 the PTUSE pulsar backend was used in parallel with the imaging observations. Details of PTUSE and how it is typically used can be found in \cite{bja+20}. In U3 we used PTUSE in pulsar search mode, this allowed us to study the single pulse behaviour of the source and thereby establish whether there was strong pulse-to-pulse modulation as well as scintillation. The observations were made with 1024 channels across the UHF band and in full polarisation mode with a sampling time of 67.7~$\mu$s. The MeerKAT B-engine was used to form the beams which were phased up and polarisation calibrated as described in \cite{bja+20}. 

We generated a series of single pulses from the PTUSE data using \texttt{DSPSR} \citep{2011vanstraten} and an ephemeris based on the best period and DM determined from the pulses detected during the U2 observation. These data were then manually cleaned of radio-frequency interference (RFI) using the \texttt{psrzap} from \texttt{PSRCHIVE} \citep{2004Hotan}. To improve the period and DM we then used \texttt{PSRCHIVE} to form an average profile for each of the 9 on-source scans made during the U3 run. Each average profile was split into 4 sub-bands of equal bandwidth in the UHF frequency range from 544--1088\,MHz. We measured arrival times for each sub-band and each scan using \texttt{pat} from  \texttt{PSRCHIVE} using an analytic template. The template was made by fitting von Mises functions to the profile with the highest S/N amongst all the scans with \texttt{paas} from \texttt{PSRCHIVE}. A fit to these arrival times using \texttt{TEMPO2} \citep{2006hobbs} results in the following best-fit parameters: period=1.63847403(2)\,s and DM=21.23(5)~pc~cm$^{-3}$ at a reference MJD epoch of 59614.0\footnote{The value in the parentheses corresponds to the 1-sigma error on the last digit.}. The DM corresponds to a distance of either 0.89\,kpc \citep[NE2001]{2002cordes} or 1.1\,kpc \citep[YMW16]{2017yao} with the usual uncertainty of between 20-30\%. 

To compare the on-pulse flux with that seen in the images we generated dynamic spectra from the PTUSE observations. As a compromise between frequency and time resolution and S/N we chose to average the data to have 256 frequency sub-bands across the UHF band, giving a frequency resolution of 2.25\,MHz and a time resolution of approximately 26\,s. Using software developed for scintillation analysis (see e.g. \cite{mac+23}) we used the average pulse profile from each of the nine 1-hour duration scans to select the on-pulse region. This region was then used to calculate the S/N for each pulse profile to generate the dynamic spectra shown in Figure \ref{fig:ptuse-ds}. The frequency sub-bands that have been removed due to the presence of RFI are shown as single-valued horizontal bands. 

The dynamic spectra are striking for their variety over the approximately 9 hours that the observations span. We can see that there is strong frequency evolution in the scintillation bandwidth and duration as well as regions of significantly enhanced flux density and all of that is seen to change throughout the observation. \revone{Inspection of the individual pulses during the bright scintillation epochs reveals that the pulsar shows modulation of factors of a few in intensity superposed on the scintillation. These changes are usually confined to a small, but different from pulse-to-pulse, phase range and only some pulses show these enhancements. We also note that there are some single pulse nulls as well, these are seen as the darker lines in the dynamic spectra (although there is some time domain RFI also present in the data). A full analysis of this single pulse variability will be undertaken in a future paper.}


\newlength{\plotwidth}
%
%

\begin{figure*}
    \setlength{\plotwidth}{0.28\textwidth}
    \begin{tabular}{@{}c@{}c@{}c@{}}%
\includegraphics[height=\plotwidth,trim={0 0 52px 0},clip]{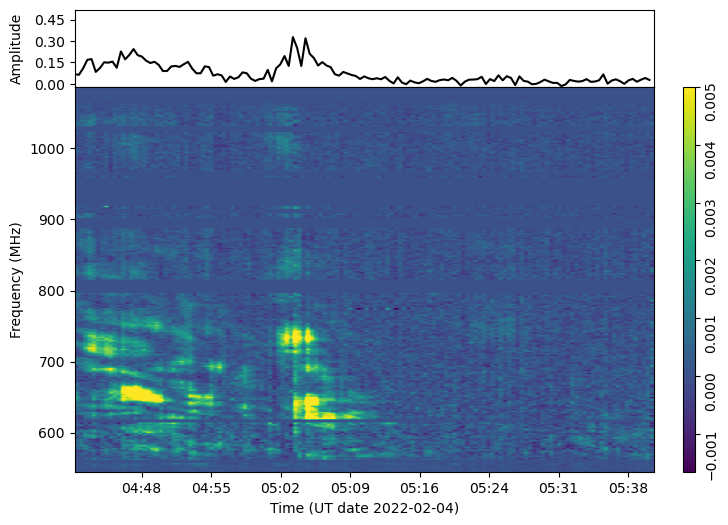}&%
\includegraphics[height=\plotwidth,trim={48px 0 52px 0},clip]{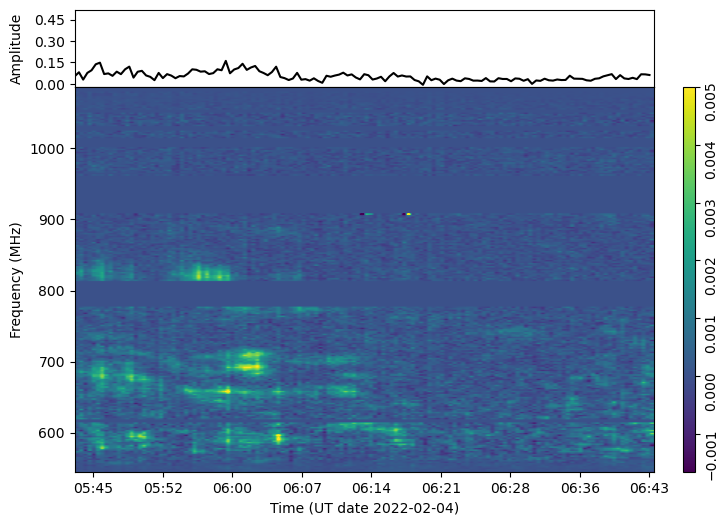}&%
\includegraphics[height=\plotwidth,trim={48px 0 0 0},clip]{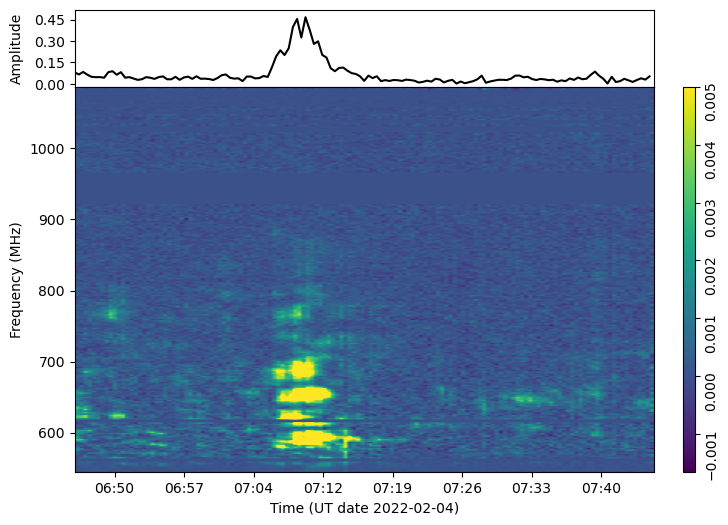}\\%
\includegraphics[height=\plotwidth,trim={0 0 52px 0},clip]{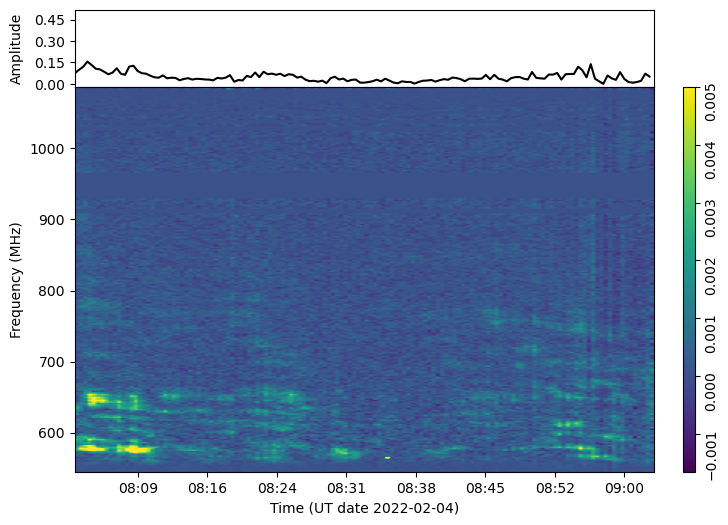}&%
\includegraphics[height=\plotwidth,trim={48px 0 52px 0},clip]{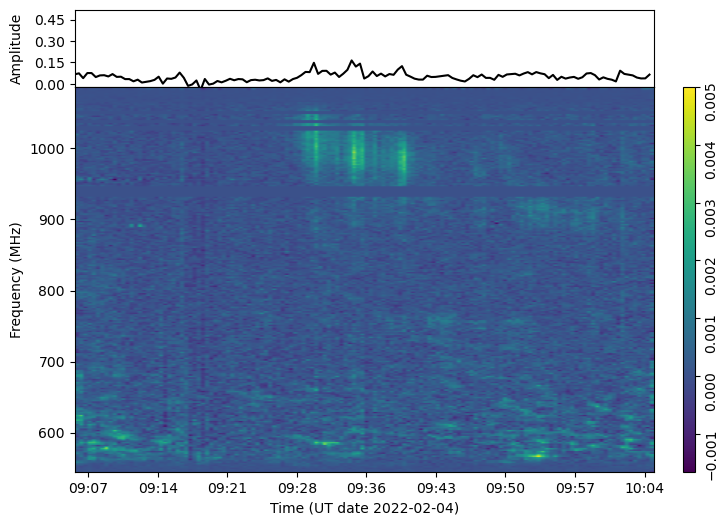}&%
\includegraphics[height=\plotwidth,trim={48px 0 0 0},clip]{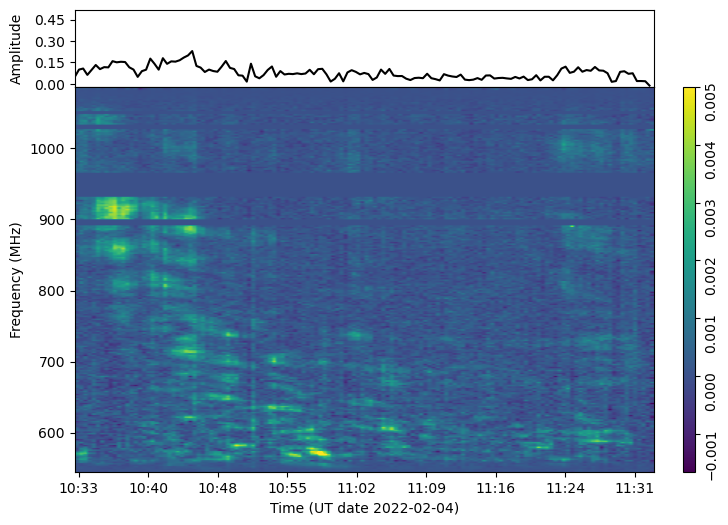}\\%
\includegraphics[height=\plotwidth,trim={0 0 52px 0},clip]{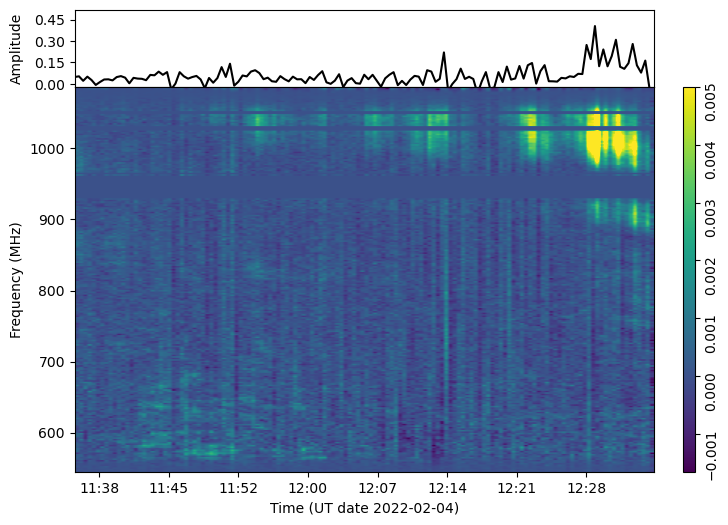}&%
\includegraphics[height=\plotwidth,trim={48px 0 52px 0},clip]{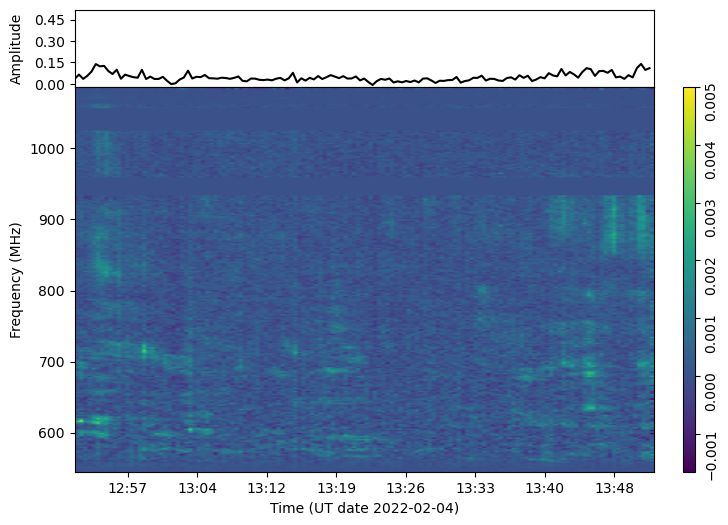}&%
\includegraphics[height=\plotwidth,trim={48px 0 0 0},clip]{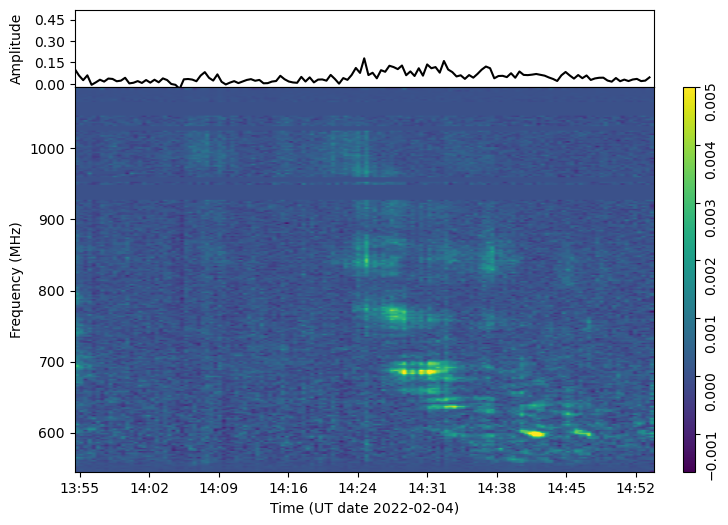}%
    \end{tabular}
    \caption{\label{fig:ptuse-ds}Dynamic spectra for the U3 observation derived from the PTUSE data showing the pulse S/N as a function of time and frequency. 
    Each plot represents one scan on the target of approximately an hour duration. The single-valued blue horizontal bands correspond to where the data has been cleaned of RFI. The data have been averaged to have a bandwidth per channel of 2.125\,MHz and each time sample corresponds to approximately 26\,s. The vertical stripes in most cases correspond to epochs when the pulsar was likely intrinsically weaker, but there is also some influence from RFI. }
\end{figure*}

\section{Discussion}
\label{sec:discuss}

The dynamic spectra obtained via PTUSE from the U3 observation (Fig.~\ref{fig:ptuse-ds}) offer the best insights into the nature of the PARROT's variability. The flux exhibits large variations over small time- and frequency-scales, and the radio-frequency scaling of the decorrelation bandwidth is similar to that observed in other radio pulsars, where the origin of the narrow-band intensity structure is understood in terms of interference arising from multi-path scattering in the interstellar medium. 
The PTUSE dynamic spectra are broadly consistent with the visibility-derived dynamic spectra (Fig.~\ref{fig:ds-others}, bottom two plots) but, unsurprisingly, show better SNR and detail. However, with the visibility-derived dynamic spectra, we can make a link to the other observations, and confirm the recurrence of similar bursts. 

In the Great Conjunction observation (Fig.~\ref{fig:ds-jup-pol}), we observe a particularly distinct ``bursting'' phase lasting about 45 minutes -- three scans -- and a ``quiescent'' phase, corresponding to all the other scans where the source is not detected in the per-scan images. However, if we make an image-plane mosaic of all the quiescent-phase scans, the source is detected with a mean flux density of 69~$\uJy$ (this corresponds to $3.5\sigma$ in a per-scan image). This is consistent with the other detections in the L-band follow-up observations (Table~\ref{tab:obs}). Presumably, the PARROT is always ``on'' and emitting pulses, which are too weak to be detected individually in the quiescent phase, but their integrated flux can be detected as a point source in the images.

The dynamic spectrum of the original detection event (Fig.~\ref{fig:ds-jup-pol}) shows a complex structure in frequency and time. The main magnification event is relatively wideband, around 500 MHz, but this is preceded and followed by a series of relatively narrower (100--150 MHz) bursts. There is also evidence of rapid (second-timescale) variability within the bursts. By contrast, all follow-up detections show bursts that are more narrowband, and fainter (Fig.~\ref{fig:ds-others}). 

\def\RMest/{$\widehat{\mathrm{RM}}$ }
\def\RM/{$\mathrm{RM}$ }

\begin{figure*}
\begin{center}
\includegraphics[width=\columnwidth]{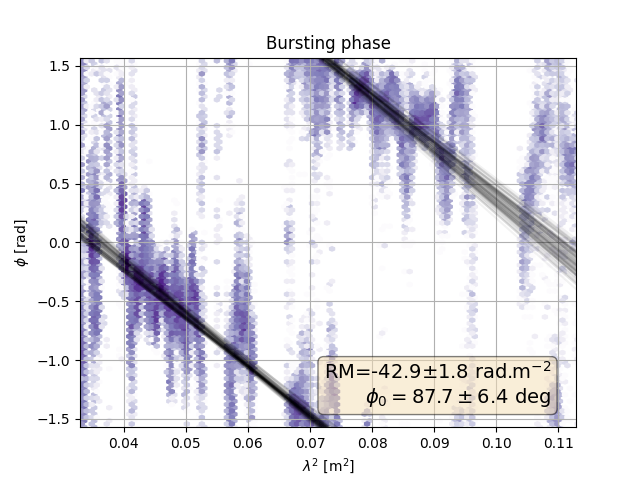}
\includegraphics[width=\columnwidth]{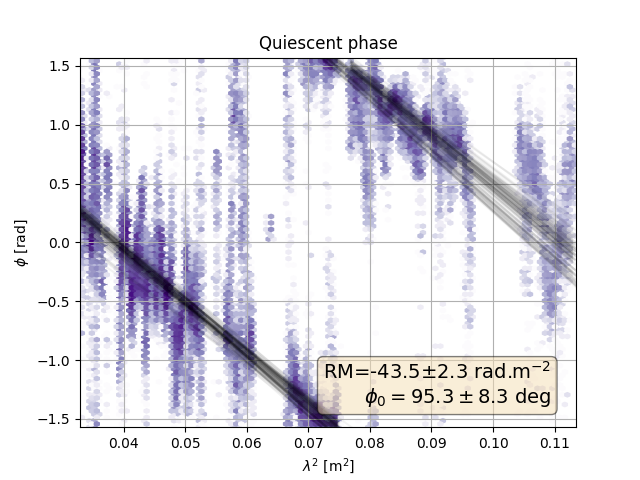}
\caption{\label{fig:RM} The left and right panels show the measured linear polarisation rotation angle as a function of $\lambda^2$ for the bursting and quiescent phases respectively. Using a single \RM/ componant model, we are able to fit the spectral behavior of the rotation angle, and find similar \RM/ values for both phases. The scattered points represent the random realisations of the rotation angle that we have used to estimate the uncertainties on \RMest/.}
\end{center}
\end{figure*}

The main detection event is almost 100\% polarized, with 89\% linear and 11\% circular polarization. Thanks to the wideband nature of the event, we are able to do a simple fit of rotation measure to the $QU$ signal (see Fig. \ref{fig:RM}). The estimated value \RMest/ of the rotation measure \RM/ is obtained by fitting a model $\phi=\mathrm{RM}\cdot\lambda^2+\phi_0$ to the measured rotation angle $\phi_m=0.5\arctan{U/Q}$. The uncertainty on \RMest/ is estimated by injecting 100 randomized samples of the noise distribution into the $QU$ dynamic spectra (the noise is estimated independently in $Q$ and $U$ using a median absolute deviation in each channel), and fitting a model to each sample. The resulting mean \RM/ of $-43\ \mathrm{rad}\,\mathrm{m}^{-2}$ in both the quiescent and bursting phases is consistent with the Galactic RM map of \citet{Galactic-RM-2020}, which gives a value of $-36 \pm 10\ \mathrm{rad}\,\mathrm{m}^{-2}$ in the direction of the PARROT.

Broadband lightcurves for the PARROT are shown in Figs.~\ref{fig:lc-rrat-1} and \ref{fig:lc-rrat-16}. The L0 lightcurve was obtained by collapsing the Stokes $I$ dynamic spectrum along the frequency axis, while for the other observations we used the image-based lightcurve extraction pipeline described in Sect.~\ref{sec:lightcurves} (the pipeline assumes an observation with a fixed
phase centre, so it was not suitable for use with the L0 observation.) 

The cause of the bursts themselves remains unclear; what we observe is very different to any established phenomenology that we are aware of. We therefore defer a thorough analysis to Paper~II, and only briefly discuss the ``suspects'' here. The two crucial features of the bursts are the characteristic timescales (minutes to tens of minutes) and the very strong brightening (almost a factor of 100 in the original detection event, and factors of 10 seen in all the follow-ups):

\begin{itemize}
\item Ionospheric scintillation is consistent with the timescales, but not with the large magnificaton factors. Furthermore, ionospheric scintillation affects sources up to a substantial angular scale \citep{TMS}, so any unusual ionospheric activitity would have imprinted a signature on the many other compact sources in the field, but we observe no such signatures (Figs.~\ref{fig:lc-u}, \ref{fig:lc-l}). We can therefore discount this explanation.
\item Interplanetary scintillation (IPS) has to be a strong suspect, since the field is in the ecliptic. Furthermore, the observation closest to the Sun (U3) did show the most bursts. However, the characteristic 
timescale of normal IPS is $\sim 1$ second \citep{TMS}; furthermore, the modulation factors of $\leq2$ \citep{zeissig-lovelace} for normal IPS are two orders of magnitude too small to account for the behaviour of our source. We note that the normal IPS phenomenology \emph{is} seen in our observations as a secondary effect. With MeerKAT's wide field-of-view (particularly in UHF) and high sensitivity, we can expect to see plenty of IPS signatures on other compact sources. This is indeed born out by Fig.~\ref{fig:lc-u}, which shows a sample of three lightcurves from the U3 observation. These sources show distinct rapid twinkling (up to 25--30\% variation) on the shortest measured timescale of 8 seconds. On the face of it then, IPS cannot explain the longer-lasting high-magification events -- these would require refractive scattering by some larger, more persistent structures in the solar wind; in turn, such structures might leave an imprint on nearby compact sources. Further analysis of this possibility is deferred to Paper II. 
\item \revone{Refractive scattering in the ISM typically manifests on far longer timescales (days to months), and with low fractional amplitudes \cite[e.g.][]{1990ARA&A..28..561R}. A small number of radio quasars display the phenomenon of intra-day variability (IDV), which is an unusually rapid form of interstellar scintillation -- see, for example, the case of J1819+3845 \citep{Dennet-Thorpe-deBruyn} and PKS 0405–385 \citep{Kedziora-Chudczer_1997} -- that is more closely matched to the timescales seen in the PARROT.  But the high magnification we observe is unprecedented for IDVs. Similarly for pulsars, some of which exhibit wide-band variations on a timescale comparable to the PARROT \cite[e.g.][]{NGC1851} --- the observed fractional amplitude of variations can be large, but nowhere near large enough to explain what we see. Indeed, we know of no instances in the astronomical literature for any type of radio source being magnified to the level displayed by the PARROT. Nor do we expect such large magnifications to arise during propagation through a turbulent medium, where the received electric field has the character of a random phasor sum. Instead it appears necessary to arrange coherent addition of the electric fields from many Fresnel zones --- smooth, refracting structures in other words.}
\item \revone{Various  types of intrinsic pulse modulation are known (giant pulses; mode-changing; nulling) that could potentially account for the wide range of fluxes that we observe from the PARROT.  However, the chromatic nature of the PARROT's bursts disfavours all intrinsic mechanisms because pulsar emission is very broad-band in character.} 
\end{itemize}

\revone{The most plausible explanation for what we see is, therefore,} some kind of unusual, strongly-refracting structure in either the ISM or the solar wind. Discriminating between these possibilities requires further analysis of the data, which we defer to a follow-up work. Note that we also have instrumental variation to contend with -- the longer-term trends in off-axis lightcurves are clearly due to azimuthal rotation of the primary beam. This is particularly apparent at L-band (Fig.~\ref{fig:lc-l}). With new primary beam measurements recently made available by \citet{mdv-beams}, these can presumably be disentangled. 

In terms of observational phenomenology then, our object \revone{seems to be neither quite a conventional pulsar, nor an RRAT}. We propose to call it a PARROT (pulsar with anomalous refraction recurring on odd timescales). \revone{The definition of what makes an RRAT varies, but we note that the fluctuations in the single pulse emission seen when the overall pulsar flux  is scintillated up, of a factor of a few, over a narrow pulse range, mean that under some conditions of scintillation, such a source could be more likely seen as an RRAT rather than a pulsar.} Most explanations for the intermittent nature of RRATs invoke intrinsic mechanisms, and assume that the pulse emission is truly ``off'' during periods of non-detection. Our imaging observations show that the PARROT is always ``on'', even when individual pulses fall below the detection threshold. However, if such anomalous refraction events are not unique, then it becomes interesting to speculate whether some known RRATs may, in fact, \revone{be PARROT-like objects}. MeerKAT's uniquely high sensitivity and ability to spatially localize such sources using imaging observations may perhaps yield new insights into the nature of RRATs and PARROTs.

What is perhaps useful to discuss at this point is how these results can inform new detection strategies for rapidly varying sources. The original PARROT discovery was the very definition of serendipity -- it is very rare for radio astronomers to make ``movies'' to begin with, and the PARROT went off in a part of the field that happened to be drawing a lot of human attention. However, its lightcurve signatures (Figs.~\ref{fig:lc-rrat-1}, \ref{fig:lc-rrat-16}) are very distinct, particularly in the L0, U2 and U3 observations, even if the narrowband magnification events are somewhat diluted by the full-band flux measurements. The dynamic spectra are even more distinct yet again. While dynamic spectra are fairly expensive to process \emph{en masse}, the lightcurve extraction pipeline described above is relatively inexpensive computationally, as well as embarrassingly parallel for most operations. It is therefore a reasonable undertaking to extract lightcurves for all sources detected in any given field routinely, with some sort of automatic detection scheme to flag up
``interesting'' lightcurves. Our results suggest that this is a very promising avenue for future discoveries. 


\section{Summary}

We have conducted full-Stokes dynamic imaging observations of the Saturn--Jupiter Great Conjunction of 2020, using the MeerKAT L-band system. Despite the complex nature of the target, we achieved an image rms of $20\mu\mathrm{Jy}$ per each 15-minute snapshot, and $7\mu\mathrm{Jy}$ in a full synthesis. The radiation belts of Jupiter were sufficiently time-variable that independent deconvolution was required per each 4-minute interval of data. We have discussed a new dynamic imaging and polarization selfcal pipeline developed specifically to process this observation.

The resolution of MeerKAT at L-band does not make Saturn a particularly compelling object of study. Jupiter's radiation belts, on the other hand, are resolved in quite some detail, and the extremely high instantaneous sensitivity of MeerKAT means that the belts' dynamics can be probed at timescales as short as minutes.

Presented in ``movie'' form, the observations revealed a transient source in close vicinity to Saturn, peaking at 5.6 mJy and remaining detectable for three scans, or 45 minutes. Dynamic spectra for the transient showed a complex frequency structure, with a series of narrowband ``bursts'' pre- and post-peak, and a fairly wideband spectrum at the peak. The source was highly polarized, with 89\% linear and 11\% circular polarization. Linear polariation showed a rotation measure consistent with the Galactic RM in the direction of the source. By imaging a subset of the data excluding the burst, we detected a compact source with a ``quiescent'' flux of $69\mu\mathrm{Jy}$.

A series of follow-up observations centred on the transient position were then carried out using the L-band and UHF systems. In all observations, we detected the source with a mean flux density of $58\sim86$ $\mu\mathrm{Jy}$ in L-band, and $102\sim231$ $\mu\mathrm{Jy}$ in UHF, with its dynamic spectra and lightcurves exhibiting occasional short-duration magnification events (``bursts'') of up to $\times10$. During two of these observations, the MeerTRAP backend detected periodic emission coincident with such a burst, which demonstrated the source to be a new radio pulsar, designated as PSR J2009$-$2026. An additional UHF-band observation was conducted commensally with the PTUSE backend which allowed us to refine the properties of the pulsar. We also determined that PSR J2009$-$2026 has a period of 1.6\,s and a DM of 21.23(5)~pc~cm$^{-3}$ and is thus at a distance of around 1\,kpc. Follow-up observations will be required to determine its period-derivative and thus locate it relative to the known pulsar population. Further study is also required to understand \revone{the details of} the pulse-to-pulse modulation properties of the source. 
\revone{PSR J2009$-$2026 shows clear signatures of scintillation in its variability, however the magnification factors it exhibits are unprecendented, and the timescales on which the scintillation pattern changes are also unusual.}

\section{Conclusions}

\revone{Our observations of Jupiter have reliably detected its radiation belts. While this has not yet yielded any new scientific insights, further analysis of the data no doubt will. It will be particularly interesting to apply the tomographic technique developed by \citet{Sault-jove} to make a 3D reconstruction of the belts, since MeerKAT provides slightly better resolution and far superior sensitivity and $uv$-coverage than the ATCA observations employed by these authors. The results of the self-calibration process discussed in Sect.~\ref{sec:jove-pipeline} make it clear that we are detecting variations in the belts on minute timescales, at high SNR, and that we have also obtained correspondingly deep polarimetry data. This is a rich vein of data awaiting further exploitation.}

\revone{We have presented a new software pipeline for detection of variable sources and transients in imaging data, with dynamic spectra and lightcurves as the data products. This showcases MeerKAT's capabilities for detecting minute-to-hour timescale transients and variables, with full-band lightcurves reliably reaching an rms below $150\mu\mathrm{Jy}$ at 8s cadence. Such sensitivity at short timescales is truly unprecedented, and promises to open up a new window on the transient and variable radio universe.}

\revone{Further analysis is required to definitively establish the nature of the observed variability of the PARROT.
The chromatic nature of the observed bursts favours some sort of strong refraction (``lensing'') mechanism, albeit with magnifications that are sometimes extraordinary. There is no reason to think that this object is unique, however, it was only detected thanks to MeerKAT's superb sensitivity. A targeted search of archival (and new) MeerKAT data can be expected to yield more such objects.}
\\
\\
\noindent{\it Acknowledgements.} The MeerKAT telescope is operated by the South African Radio Astronomy Observatory, which is a facility of the National Research Foundation, an agency of the Department of Science and Innovation. OMS's research is supported by the South African Research Chairs Initiative of the Department of Science and Technology and National Research Foundation (grant No. 81737). BWS, MC, KMR and MeerTRAP acknowledge funding from the European Research Council (ERC) under the European Union’s Horizon 2020 research and innovation programme (grant agreement No. 694745). PAW acknowledges financial support from the National Research Foundation and the University of Cape Town. LR appreciates the support from the Swedish National Space Agency through grant 2021-00153. We thank Andrew Jameson for assistance with the use of PTUSE and the data reduction. We thank Robert Main for providing the software that was used to produce the dynamic spectra from the PTUSE data. We thank Ana Corrochano Lopez for inspiring the original conjunction observation.
\\
\\
\noindent{\it Data Availability.} The raw data underlying this article is publicly available via the SARAO archive\footnote{\url{https://archive.sarao.ac.za}}, under proposal ID SSV-20200715-SA-01. Additional data products, such as 
the conjunction movie, are available via \url{https://doi.org/10.48479/shtp-2015}. Processing recipes and scripts are available via \url{https://ratt.center/parrot}.

\bibliographystyle{mnras}
\bibliography{references}
\label{lastpage}
\end{document}